\documentclass[twocolumn]{aastex63}
\usepackage{graphicx}
\usepackage{lineno}
\usepackage{multirow,tabularx}

\def\iso#1{$^{#1}$}
\def\msun{M$_\odot$}

\submitjournal{ApJ}

\shorttitle{}
\shortauthors{Trueman et al.}
\graphicspath{{./}{figures/}}
\begin{document}

\title{Galactic Chemical Evolution of Radioactive Isotopes with an \textit{s}-process Contribution}

\correspondingauthor{Thomas Trueman}
\email{thomas.trueman@csfk.mta.hu}

\author{Thomas C. L. Trueman}
\affiliation{Konkoly Observatory, Research Centre for Astronomy and Earth Sciences, Eötvös Loránd Research Network, Konkoly Thege Mikl\'{o}s \'{u}t 15-17, H-1121 Budapest, Hungary}
\affiliation{E.A. Milne Centre for Astrophysics, Department of Physics \& Mathematics, University of Hull, HU6 7RX, UK}
\affiliation{NuGrid Collaboration, \url{http://nugridstars.org}}

\author{Benoit Côté}
\affiliation{Konkoly Observatory, Research Centre for Astronomy and Earth Sciences, Eötvös Loránd Research Network, Konkoly Thege Mikl\'{o}s \'{u}t 15-17, H-1121 Budapest, Hungary}
\affiliation{ELTE E\"{o}tv\"{o}s Lor\'{a}nd University, Institute of Physics, Budapest 1117, P\'{a}zm\'{a}ny P\'{e}ter s\'{e}t\'{a}ny 1/A, Hungary}
\affiliation{NuGrid Collaboration, \url{http://nugridstars.org}}
\affiliation{Joint Institute for Nuclear Astrophysics - Center for the Evolution of the Elements, USA}

\author{Andr\'es Yag\"ue L\'opez}
\affiliation{Konkoly Observatory, Research Centre for Astronomy and Earth Sciences, Eötvös Loránd Research Network, Konkoly Thege Mikl\'{o}s \'{u}t 15-17, H-1121 Budapest, Hungary}
\affiliation{NuGrid Collaboration, \url{http://nugridstars.org}}

\author{Jacqueline den Hartogh}
\affiliation{Konkoly Observatory, Research Centre for Astronomy and Earth Sciences, Eötvös Loránd Research Network, Konkoly Thege Mikl\'{o}s \'{u}t 15-17, H-1121 Budapest, Hungary}
\affiliation{NuGrid Collaboration, \url{http://nugridstars.org}}

\author{Marco Pignatari}
\affiliation{Konkoly Observatory, Research Centre for Astronomy and Earth Sciences, Eötvös Loránd Research Network, Konkoly Thege Mikl\'{o}s \'{u}t 15-17, H-1121 Budapest, Hungary}
\affiliation{E.A. Milne Centre for Astrophysics, Department of Physics \& Mathematics, University of Hull, HU6 7RX, UK}
\affiliation{NuGrid Collaboration, \url{http://nugridstars.org}}
\affiliation{Joint Institute for Nuclear Astrophysics - Center for the Evolution of the Elements, USA}

\author{Benj\'{a}min Soós}
\affiliation{Konkoly Observatory, Research Centre for Astronomy and Earth Sciences, Eötvös Loránd Research Network, Konkoly Thege Mikl\'{o}s \'{u}t 15-17, H-1121 Budapest, Hungary}
\affiliation{ELTE E\"{o}tv\"{o}s Lor\'{a}nd University, Institute of Physics, Budapest 1117, P\'{a}zm\'{a}ny P\'{e}ter s\'{e}t\'{a}ny 1/A, Hungary}

\author{Amanda I. Karakas}
\affiliation{School of Physics and Astronomy, Monash University, VIC 3800, Australia}
\affiliation{ARC Centre of Excellence for All Sky Astrophysics in 3 Dimensions (ASTRO 3D)}

\author{Maria Lugaro}
\affiliation{Konkoly Observatory, Research Centre for Astronomy and Earth Sciences, Eötvös Loránd Research Network, Konkoly Thege Mikl\'{o}s \'{u}t 15-17, H-1121 Budapest, Hungary}
\affiliation{ELTE E\"{o}tv\"{o}s Lor\'{a}nd University, Institute of Physics, Budapest 1117, P\'{a}zm\'{a}ny P\'{e}ter s\'{e}t\'{a}ny 1/A, Hungary}
\affiliation{School of Physics and Astronomy, Monash University, VIC 3800, Australia}

\begin{abstract}

Analysis of inclusions in primitive meteorites reveals that several short-lived radionuclides (SLRs) with half-lives $0.1-100$ Myr existed in the early Solar System (ESS). We investigate the ESS origin of $^{107}$Pd, $^{135}$Cs, and $^{182}$Hf, which are produced by $slow$ neutron captures (the $s$-process) in asymptotic giant branch (AGB) stars. We modelled the galactic abundances of these SLRs using the \texttt{OMEGA+} galactic chemical evolution (GCE) code and two sets of mass- and metallicity-dependent AGB nucleosynthesis yields (Monash and FRUITY). Depending on the ratio of the mean life $\tau$ of the SLR to the average length of time between the formation of AGB progenitor $\gamma$, we calculate timescales relevant for the birth of the Sun. If $\tau/\gamma\gtrsim2$, we predict self-consistent isolation times between 9 and 26 Myr by decaying the GCE predicted $^{107}$Pd/$^{108}$Pd, $^{135}$Cs/$^{133}$Cs, and $^{182}$Hf/$^{180}$Hf ratios to their respective ESS ratios. The predicted $^{107}$Pd/$^{182}$Hf ratio indicates that our GCE models are missing $9-73\%$ of $^{107}$Pd and $^{108}$Pd in the ESS. This missing component may have come from AGB stars of higher metallicity than those that contributed to the ESS in our GCE code. 
If $\tau/\gamma\lesssim0.3$, we calculate instead the time ($T_{\rm LE}$) from the last nucleosynthesis event that added the SLRs into the presolar matter to the formation of the oldest solids in the ESS. For the 2 \msun, $Z=0.01$ Monash model we find a self-consistent solution of $T_{\rm LE}=25.5$ Myr. 

\end{abstract}


\section{Introduction} \label{sec:intro}

It has been inferred from analysis of meteoritic rocks and inclusions that many short-lived radionuclides (SLRs) with half-lives of $T_{1/2}\sim 0.1-100$ Myr were present in the early solar system \citep[ESS;][]{dauphas, lugaro18}. These SLRs can be used as tracers of the local circumstances of the birth of the Sun and the history that led to them. In particular, SLRs offer the unique opportunity to probe the length of time that the protosolar gas was isolated from further stellar enrichment events in the Galaxy before the birth of the Sun - the so called, ``isolation time" \citep[see e.g.][]{wasserburg06, huss09,lugaro14,cote19}. Furthermore, the in situ decay of $^{26}$Al ($T_{1/2}=0.72$ Myr) provided an important energy source for the thermo-mechanical evolution of protoplanets \citep[see, e.g.,][]{lichtenberg}. However, a self-consistent origin scenario that explains the abundances of of all SLRs in the ESS has yet to be found.

We investigate the ESS origin of three SLRs\footnote{Despite the fact it is also made by the \textit{s}-process, we do not study $^{205}$Pb due to its poorly understood half-life in stellar interiors \citep{mow98}. }, $^{107}$Pd ($T_{1/2}=6.5$ Myr), $^{135}$Cs ($T_{1/2}= 2.3$ Myr), and $^{182}$Hf ($T_{1/2}= 8.90$ Myr), which can be produced in astrophysical sites by two neutron capture processes \citep{burbidge57}: the \textit{slow} (\textit{s})-process \citep[see review by][]{kap11}, so called because the neutron capture rate is slow compared to the $\beta$-decay rate of unstable nuclei along the $s$-process path; and the \textit{rapid} (\textit{r})-process \citep[see review by][]{thi11}, where instead the timescale for neutron-capture is much shorter than the competing $\beta$-decay rate. The main site of production of the \textit{s}-process isotopes in the mass range $90<\text{A}<208$ are low- and intermediate-mass ($M\lesssim8$ \msun) asymptotic giant branch (AGB) stars \citep{gal98, busso99, gor00, her05, cristallo09, lugaro12}. These stars experience thermal convective instabilities triggered by recurrent He-burning episodes on top of a degenerate C-O core. During the relatively long interpulse phase ($\gtrsim10^{3}$ years), the $^{13}\text{C}(\alpha,n)^{16}\text{O}$ neutron source reaction is activated. The nucleosynthesis products are subsequently mixed into the thermal convective instability region where a second neutron source, the $^{22}$Ne$(\alpha,n)^{25}$Mg reaction, is marginally activated and plays a role in the production of several $s$-process isotopes. These products are then brought to the surface of the star in a convective mixing process called third dredge-up (TDU), from where they are then lost to the interstellar medium (ISM) via stellar winds \citep[][and references therein]{karakas2014}. 

Two scenarios have been proposed to explain the origin of $^{107}$Pd, $^{135}$Cs, and $^{182}$Hf in the ESS: $(1)$ a nearby star that ejected its material into the protosolar nebula, and $(2)$ Galactic inheritance from the local ISM. Considering first scenario $(1)$, the idea that a nearby, low-mass AGB could have polluted the ESS was explored by \cite{wasserburg94} and \cite{busso99}. The former found that such stars can readily provide a solution for the origin of several SLRs in the ESS, and that the production of $s$-isotopes is sensitive to the overall neutron exposure. However, as \cite{busso99} point out, the AGB models used in \cite{wasserburg94} incorrectly assumed that the $^{13}\text{C}(\alpha,n)^{16}\text{O}$ reaction takes place under convective rather than radiative conditions. More recent studies have looked in to an intermediate-mass AGB star as a potential candidate \citep{wasserburg06,wasserburg17,vesconi18}, however, these stars produce, for example, too much $^{107}$Pd relative to $^{26}$Al to provide a self consistent solution. Core-collapse supernovae stars have also been considered as potential sources for the origin of $^{107}$Pd and $^{182}$Hf \citep[][Lawson et al. in prep]{meyer00}. However, a late addition of ejecta from a massive star into the forming solar nebula would significantly overproduce $^{53}$Mn and $^{60}$Fe if the dilution factor is calibrated to reproduce the necessary abundances for other SLRs \citep[see, e.g.][]{wasserburg06,vesconi18}. 

In scenario $(2)$ the distribution of SLRs in the ESS reflects contributions from multiple stellar enrichment sources. In this case, the abundance of isotopes in the ISM at the time of the birth of the Sun ($t_{\odot}$) can be predicted using a galactic chemical evolution (GCE) code that considers contributions from a variety of stellar nucleosynthesis events \citep[][]{trav99, trav04, huss09, pra18, pra20, cote19}. To compare to the ESS data, we need to calculate the ratio of an SLR relative to a stable, or long-lived, reference isotope, which directly probes the complete star formation and gas flow histories of the Milky Way. The isolation time ($T_{\text{iso}}$) is therefore the time taken for the radioactive-to-stable abundance ratio in the ISM at $t_{\odot}$ to reach the abundance ratio inferred for the earliest solids known to form in the ESS, assuming that the only change in the relative abundances of the two isotopes is due to the radioactive decay of the SLR. 

The uncertainties associated with the modelling of the evolution of radioactive-to-stable isotopic ratios in the Galaxy were analysed quantitatively by \cite{cote19} (hereafter Paper I). In Paper I the GCE framework used herein was first established: using observations of the Galactic disk to calibrate a two-zone GCE code, \texttt{OMEGA+}, a low, a best, and a high value of the radioactive-to-stable isotopic ratio in the ISM at $t_{\odot}$ were obtained with a total uncertainty of a factor of $3.6$ between the high and low GCE setups. To quantify the uncertainty in the isotopic ratio, considering the fact that enrichment events are not continuous but discrete in time, \cite{cote19b} (hereafter Paper II) added Monte Carlo calculations to the GCE framework in order to sample appropriate delay-time distribution functions for different astrophysical sites. They recovered uncertainty factors for the abundance of an SLR in a given parcel of ISM matter at $t_{\odot}$ for several values of $\tau/\gamma$, where $\tau$ is the mean life of the SLR ($\tau = T_{1/2}/\ln2$) and $\gamma$ is the interval of time between the formation of enrichment progenitor. If $\tau/\gamma\gtrsim2$ (henceforth Regime I), the GCE description is valid and the error bar of the abundance of the SLR at $t_{\odot}$ remains below a factor of 0.8. If $\tau/\gamma\lesssim0.3$ (henceforth Regime III), instead the SLR probably originates from only one event, so it is not possible to determine the isolation time but only the time from the last event \citep[see discussion in][]{lugaro14, lugaro18}. \cite{yague21} (hereafter Paper III) expanded this framework to analyse the ratio of two SLRs that are produced together by the same enrichment events. By exploring this last ratio, one can completely remove uncertainties associated with the GCE of the stable isotopes whilst simultaneously reducing the statistical uncertainty on the radioactive-to-radioactive ratio due to ISM heterogeneities to less than a factor of a half, assuming that Regime I holds for both SLRs.        

The aim of this paper is to analyse $^{107}$Pd, $^{135}$Cs, and $^{182}$Hf together in the framework described above. It is possible to do this because \cite{lugaro14} showed that $^{182}$Hf has a substantial \textit{s}-process component in the Galaxy, since the faster decay of $^{181}$Hf at stellar temperatures, due to the existence of a 68 keV excited level and hampering $^{182}$Hf production, was based on a wrong assignment \citep{rickey68}, as already noted in \cite{firestone91}, and then confirmed by \cite{bon02}. Prior to this, it was found that the majority of $^{182}$Hf, like $^{129}$I ($T_{1/2}=15.7$ Myr), was mostly produced by the \textit{r}-process leading to inconsistent isolation times from $^{182}$Hf/$^{180}$Hf and $^{129}$I/$^{127}$I \citep[see, e.g.,][and references therein]{ott08}. With the knowledge now that $^{182}$Hf is also produced by the \textit{s}-process in AGB stars, we can use the new methodology and framework presented in Papers I and II to re-investigate the SLRs with an \textit{s}-process Galactic component. 

\section{Methods and models} \label{sec:methods}

Following the methodology of Paper I, we calibrate our GCE framework to recover a low, best, and high value for the radioactive-to-stable abundance ratio of $^{107}$Pd/$^{108}$Pd, $^{135}$Cs/$^{133}$Cs, and $^{182}$Hf/$^{180}$Hf in the ISM at $t_{\odot}$. Paper I only considered the simplified case of a constant production ratio between an SLR and its reference isotope, while in this work we use mass- and metallicity-dependent stellar nucleosynthesis yields from two different sets of AGB models in our GCE framework. In addition, we use the uncertainties calculated in Papers II and III to account for the effects of ISM heterogeneities on the radioactive-to-stable and radioactive-to-radioactive abundance ratios at $t_{\odot}$. In this Section we give a brief description of our GCE framework, the choice of stellar yields, and the calibration of our three different GCE setups.

\subsection{The \texttt{OMEGA+} GCE framework}\label{sec:omega}

We follow the evolution of SLRs in the Galaxy using the publicly available \texttt{OMEGA+} GCE code\footnote{https://github.com/becot85/JINAPyCEE} \citep{cote18}. This two-zone model is comprised of $(1)$ a central star forming region, modelled using the \texttt{OMEGA} code \citep[]{cote17}, which simulates the chemical evolution of a cold gas reservoir as a function of time, and $(2)$ a surrounding hot gas reservoir with no star formation. Following the nomenclature adopted in \cite{cote18}, we refer to Region $(1)$ as the galaxy and Region $(2)$ as the circumgalactic medium (CGM).

At each timestep, the code creates a simple stellar population in the galaxy. All stars in a stellar population form at the same time and have the same initial metallicity - that of the ISM at that time - since they are assumed to have been born from the same parent gas cloud. The mass of the stellar population is proportional to the star formation rate (SFR) at that time. The SFR at any given time is directly proportional to the total mass of gas inside the galaxy, such that

\begin{equation}
    \dot{M}_{\star}(t) = \frac{\epsilon_{\star}}{\tau_{\star}}M_{\text{gas}}=f_{\star}M_{\text{gas}}(t),
\end{equation}
where $f_{\star}$ [yr$^{-1}$], represents the combination of the dimensionless star formation efficiency $\epsilon_{\star}$ and the star formation timescale $\tau_{\star}$. At each time step in the simulation, the \texttt{SYGMA} code \citep[][]{rit18} calculates the combined yield from all stellar populations in the galaxy. For a galaxy with \textit{N} stellar populations formed by time \textit{t}, each with their own initial mass, metallicity, and formation time ($M_j$, $Z_j$, and $t_j$ respectively), the rate at which gas is returned to the ISM from stellar ejecta is given by

\begin{equation}
    \dot{M}_{\text{ej}}(t) = \sum_j{\dot{M}_{\text{ej}}(M_j, Z_j, t-t_j)}.
\end{equation}
where $t-t_{j}$ is the current age of population $j$. 
The code includes the mass- and metallicity-dependent yields for low- and intermediate-mass stars (the progenitors of AGB stars), massive stars and their core-collapse supernovae, and SNe Ia. Additionally, it has the option to include contributions from any number of user-defined additional sources. \texttt{OMEGA} is a one-zone GCE model, so the stellar ejecta is assumed to mix instantenously and uniformly into the ISM.

The addition of the CGM as a one-zone extension to \texttt{OMEGA} allows the code to track the elements that are expelled from the galaxy by galactic outflows. Whilst galactic inflows introduce new, often metal poor gas into the galaxy, galactic outflows expel gas and heavy elements into the CGM. Considering the transfer of matter into and out of the galaxy, the gas locked inside stars ($M_{\star}$), and the gas returned to the ISM by stellar ejecta ($M_{\text{ej}}$), the time dependence of the total mass ($M_{\text{gas}}$) inside the galaxy can be expressed as

\begin{equation}
    \dot{M}_{\text{gas}}(t) = \dot{M}_{\text{inflow}}(t) + \dot{M}_{\text{ej}}(t) - \dot{M}_{\star}(t) - \dot{M}_{\text{outflow}}(t).
\end{equation}

The gas inflow and outflow rates can be controlled by user-defined inputs. In this work we apply the methodology from \cite{chi97} and assume two exponential gas inflow episodes described by 

\begin{equation}
    \dot{M}_{\text {inflow }}(t)=A_{1} \exp \left(\frac{-t}{\tau_{1}}\right)+A_{2} \exp \left(\frac{t_{\max }-t}{\tau_{2}}\right),
\end{equation}
where $A_1$ and $A_2$ are the normalization of the first and second infall, $\tau_{1}$ and $\tau_{2}$ are the timescales for mass accretion in the first and second infall episodes, and $t_{\text{max}}$ is the time of maximum contribution of the second gas accretion episode which is assumed to be zero for the first episode. For all GCE setups, $\tau_{1}=0.68$ Gyr, $\tau_{2}=7.0$ Gyr and $t_{\text{max}}=1.0$ Gyr. The outflow rate is proportional to the SFR,

\begin{equation}
    \dot{M}_{\text {outflow }}(t)=\eta \dot{M}_{\star}(t),
\end{equation}
where the the mass-loading factor, $\eta$, determines the magnitude of the outflow.  

\subsection{Stellar AGB yields} \label{sec:yields}

\begin{figure*}
    \centering
    \includegraphics[scale=0.9]{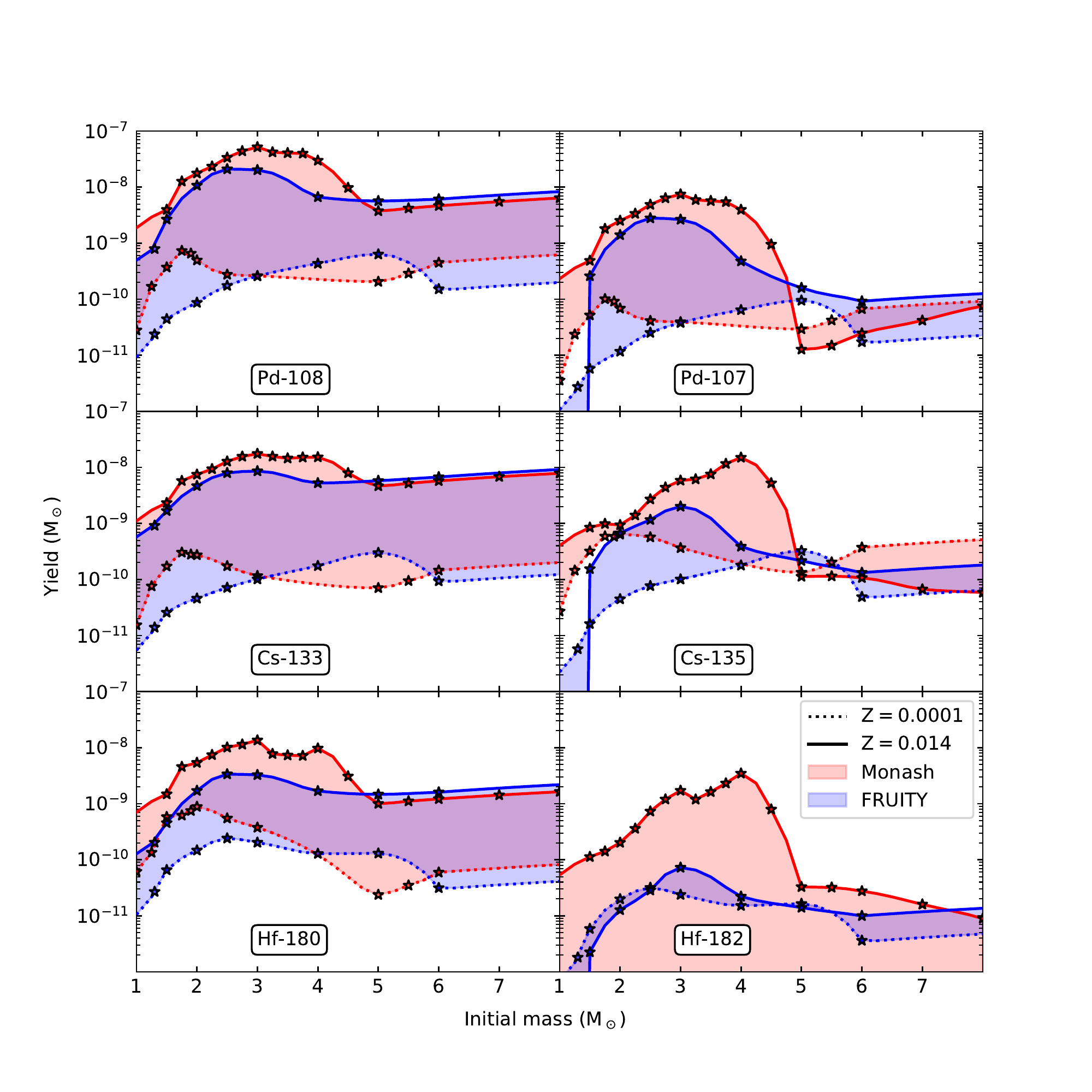} 
    \caption{Predicted isotopic yields from the Monash (red) and FRUITY (blue) AGB stellar nucleosynthesis models. The yields from models of two metallicities ($0.0001$ and $0.014$) are plotted as lines and star symbols as a function of initial stellar mass for the radioactive isotopes $^{107}$Pd, $^{135}$Cs, and $^{182}$Hf, as well as their respective stable reference isotope. The shaded regions indicate the range of yields ejected by models with $0.0001<Z<0.014$. Yields with an initial mass marked by a star symbol are taken directly from a model, while the lines are obtained by their interpolation.}
    \label{fig:yields}
\end{figure*}

Low- and intermediate-mass stars ($\sim0.8-8$ \msun) will evolve along the AGB stage of evolution after core H- and He-burning. Structurally, AGB stars are characterized by a degenerate C/O core surrounded by an inner He-burning shell, which is separated from an outer H-burning shell by a He-rich intershell region. The outermost layer of the star is an extended H-rich convective envelope which experiences mass loss via stellar winds. In the advanced stage of AGB evolution, the star undergoes recurrent He-burning flashes (or thermal pulses, TPs), at the base of the intershell \citep{her05, karakas2014}. Each TP releases a large amount of energy for a short period of time, which drives a convective region over the whole intershell. As the star expands, the temperature in the H-burning shell falls below that required to sustain nuclear fusion, and H burning switches off. Once the TP is extinguished, and before H burning starts again, the outer base of the convective envelope can reach down deeper into the star. The ashes of He burning can then be mixed into the envelope and brought to the surface, a process called third dredge-up (TDU). It is during this thermally-pulsing AGB phase that the two neutron source reactions, $^{13}\text{C}(\alpha,n)^{16}\text{O}$ and $^{22}\text{Ne}(\alpha,n)^{25}\text{Mg}$, are activated.  

The $^{13}\text{C}(\alpha,n)^{16}\text{O}$ reaction dominates the production of the $s$-process isotopes in low-mass AGB stars and is activated during the periods of quiescent H-burning between each TP. Although the neutron densities rarely exceed $10^7\text{cm}^{-3}$, the long timescales during the interpulse periods ($\sim10^4$ years) means that the overall neutron exposure (essentially the neutron density integrated over time) is high. In order for enough neutrons to be released via the $^{12}\text{C}(p,\gamma)^{13}\text{N}(\beta^+)^{13}\text{C}$ reaction chain, at the deepest extent of the TDU a partial mixing zone (PMZ) forms where protons are mixed into the intershell from the convective envelope. This results in the formation of a thin $^{13}$C-rich region at the top of the intershell, called the $^{13}$C ``pocket". The formation of the PMZ is a long standing uncertainty in AGB nucleosynthesis models and a common consensus regarding its implementation in 1D stellar evolution models has yet to be reached \citep[see][for further discussion]{wag2020}. Proposed mixing mechanisms include, but are not limited to, diffusive mixing \citep{her97, cristallo09}, internal gravity waves \citep{den03}, convective boundary mixing \citep{bat16}, and magneto driven hydrodynamics \citep{trip16, bus20}. Nevertheless, \cite{bun17} demonstrated that the nature of the mixing function that creates the pocket in 1-D stellar models, once the size of the PMZ is defined, is generally a smaller source of uncertainty on the $s$-process nucleosynthesis than other uncertainties in stellar physics, like the treatment of convective boundaries. 

During a TP, the $^{14}$N in the H ashes is entirely consumed to make $^{22}$Ne by the reaction chain $^{14}\text{N}(\alpha,\gamma)^{18}\text{F}(\beta,\nu)^{18}\text{O}(\alpha,\gamma)^{22}\text{Ne}$. In AGB stars with an initial mass $\gtrsim3$ \msun\ the temperatures are high enough during a TP to activate the $^{22}\text{Ne}(\alpha,n)^{25}\text{Mg}$ neutron source reaction. The resulting $s$-process takes place over a much shorter timescale ($\approx10$ years), but with higher neutron densities (up to $\sim10^{11}\text{cm}^{-3}$) than from the $^{13}\text{C}(\alpha,n)^{16}\text{O}$ neutron source. Overall, the time integrated neutron flux is lower from this secondary neutron burst, which prevents the production of $s$-nuclei beyond the Sr-peak\footnote{The solar \textit{s}-process abundance distribution has three abundance peaks, which arise due to lower neutron capture cross-sections at the neutron magic numbers $N=50,82,126$.}. However, the $^{22}\text{Ne}(\alpha,n)^{25}\text{Mg}$ plays a crucial role in the activation of branching points at unstable nuclei along the $s$-process path \citep{bis15}. Branching point nuclei are so called because their rate of $\beta$-decay is comparable to the rate of neutron capture, thus the $s$-process can \textit{branch} in two different ways. Of relevance here, are the branching points at $^{134}$Cs ($T_{1/2}=2.1$ years) and $^{181}$Hf ($T_{1/2}=42.3$ days), which lead to the production of $^{135}$Cs and $^{182}$Hf, respectively. The production of $^{135}$Cs and $^{182}$Hf therefore require the activation of both neutron sources: first, the $^{13}\text{C}(\alpha,n)^{16}\text{O}$ reaction is needed to produce the stable isotopes $^{133}$Cs and $^{180}$Hf, then the $^{22}\text{Ne}(\alpha,n)^{25}\text{Mg}$ reaction is needed to activate the $^{134}$Cs and $^{181}$Hf branching points. This condition is only met in AGB stars with initial mass $\sim3-4$ \msun, in which both neutron sources are activated relatively efficiently. AGB stars of lower masses do not activate the $^{22}\text{Ne}$ neutron source, while at higher masses the mixing leading to the formation of the $^{13}$C pocket is inhibited by the pressure and density distribution in the very thin intershell \citep{cris11}, and the hot temperature at the base of the convective envelope at the time of the TDU \citep{gor04}.

We model the GCE of the \textit{s}-process SLRs using two sets of AGB nucleosynthesis yields: $(1)$ Monash \citep{kar12,fish14,kar16,kar18} and $(2)$ FRUITY \citep{cris08, cristallo09, cris11, cris15}. We also considered yields from the NuGrid \citep{nugrid1, nugrid2, bat19, bat21} and the S-process NUcleosynthesis Post-Processing code for ATon \citep[\texttt{SNUPPAT;}][accepted]{snuppat} set of AGB models. However, the range of masses and metallicities are more limited for these data sets, and as such the results are within the variations for Monash and FRUITY GCE setups. We therefore do not use these yields in our analysis. 

For Monash, we included the mass- and metallicity-dependent yields for 82 stellar nucleosynthesis models, covering a mass range of $1.0$ \msun\ $\leq M\leq 8$ \msun\ for metallicities $0.0001\leq Z \leq 0.03$. In these models the PMZ mixing profile is artificially inserted as an exponential profile and the mass of the $^{13}$C pocket can be varied \citep[for more details see][]{bun17}. Where several models with a different size of the PMZ are available with the same initial mass and metallicity, we selected the model using the "standard" PMZ size as defined in \cite{kar16}. 

For FRUITY, we included the yields from 82 stellar nucleosynthesis models, for masses and metallicities in the ranges $1.0$ \msun\ $\leq M\leq 6$ \msun\ and $0.0001\leq Z \leq 0.02$, respectively. To populate the FRUITY yield table for stars between 6 and 8 \msun\ we use the same abundance pattern as for the highest mass model available (usually 6 \msun) at the desired metallicity. However, the total ejected mass for a star with initial mass between 6 and 8 \msun\ is found by linearly extrapolating the ejected mass as a function of initial mass for all stellar models with the same initial metallicity. We only use the non-rotating models, as the effect of rotation is likely overestimated \citep{cseh18, denhartogh}. For stars $M_{\star}>12$ \msun, we used the yields from the NuGrid massive star models \citep{nugrid2}. To obtain the yields for stars $8>M_{\star}>12$ we apply the same technique used for estimating the $6-8$ \msun\ FRUITY yields, but we instead use the yields of the 12 \msun\ and re-scale them according to a linear fit of the total ejected mass as a function of stellar mass for the NuGrid massive star models. Massive stars eject large amounts of the $A<90$ $s$-elements into the ISM following the activation of the $^{22}\text{Ne}(\alpha,n)^{25}\text{Mg}$ neutron source during He- and C-burning \citep[see e.g.,][and references therein]{pignatari:2010}. However, they produce only very small amounts of the SLRs of interest here.

Monash and FRUITY use different methods to create the $^{13}$C pocket: artificial injection of protons into the intershell following each TDU in the Monash models and time-dependent convective overshoot in the FRUITY models. Furthermore, the nuclear physics is different between the two yield sets, most notably for the SLRs of interest in this work is the choice of the reaction rate for the $^{181}$Hf $\beta$-decay. FRUITY uses the FU11-Network (FUN) stellar evolution code \citep{str06}, for which $\beta$-decay rates are taken from \cite{tak87}, Monash instead adopts the $\beta$-decay rate from \cite{lugaro14} based on the experimental data of \cite{bon02}. The former includes a 68 keV excited level in $^{181}$Hf, which significantly reduces the half-life in stellar interiors. The existence of this level was based on the wrong assignment of an observed band head in (d,p) by \cite{rickey68}. This was superseded by \cite{bon02}, who found no evidence of such a state\footnote{See also the latest two releases of the Nuclear Data Sheets for A=181 \citep{firestone91, wu05}.}. The absence of this state essentially removed the temperature dependence of the half-life. Using this updated reaction rate results in a significantly increased production of $^{182}$Hf in AGB nucleosynthesis models \citep{lugaro14}. 

Figure \ref{fig:yields} shows the mass- and metallicity-dependent yields of Monash and FRUITY for the SLRs $^{107}$Pd, $^{135}$Cs, and $^{182}$Hf, and their respective stable reference isotopes, $^{108}$Pd, $^{133}$Cs, and $^{180}$Hf. The stellar yields represent the total mass of a given isotope ejected over the complete lifetime of the star. The stellar yields for two metallicities ($0.0001$ and $0.014$) are plotted with lines and star symbols because they represent the extreme values for the initial compositions for which both Monash and FRUITY have yields available; the shaded regions indicate the range of yields ejected by the respective set of models between these two metallicities. We note that Monash has calculated yields for models with initial metallicity $0.03$ which are not shown in Figure \ref{fig:yields}, however, in our GCE framework stars with $Z>Z_{\odot}=0.014$ \citep{asp09} are only born after $t_{\odot}$, so they do not contribute to the ESS SLR abundances. This is in contrast with observations of the solar neighbourhood age-metallicity distribution, which shows that stars older than the Sun with $Z>Z_{\odot}$ do exist, up to [Fe/H]$\sim+0.5$. Therefore AGB stars with these metallicities may in fact have contributed to the SLR budget in the ESS and we will discuss this possibility in Section \ref{sec:LE}.

As previously mentioned, the $^{181}$Hf decay rate is different in Monash and in FRUITY: the longer half-life of $^{181}$Hf in the Monash models means it is more likely for $^{181}$Hf to capture a neutron before decaying than in the FRUITY models. This is why there is a higher production of $^{182}$Hf at Z$=0.014$ in the Monash models, up to two orders of magnitude higher than FRUITY at $M\sim 4$  \msun. We note that the lowest metallicity Monash models do not include any $^{182}$Hf\footnote{The $Z=0.0001$ models were published prior to the \cite{lugaro14} paper, and $^{181}$Hf was not included in the network.}, which is why there is no lower limit on the yields in this case. This has little impact on the final abundance of the isotope in the ESS because in our GCE framework low-metallicity stars are only born in the early Galaxy and subsequently all of the $^{182}$Hf they may produce will have decayed by $t_{\odot}$.   

A comparison of the yields from the 3 \msun, $Z=0.014$ Monash and FRUITY models for the SLRs of interest and their respective reference isotopes is shown in Table \ref{tab:yields}. For sake of comparison, a 3 \msun\ model is chosen since it represents the typical mass at which both the $^{13}\text{C}(\alpha,n)^{16}\text{O}$ and the $^{22}\text{Ne}(\alpha,n)^{25}\text{Mg}$ neutron source reactions are activated.

Comparing the absolute yields in Table \ref{tab:yields}, we can see that the Monash yields are higher for all the isotopes. In the case of the $^{107}$Pd/$^{108}$Pd ratio, the models show good agreement. This is because the $^{107}$Pd/$^{108}$Pd ratio is mostly determined by the inverse ratio of the neutron-capture cross sections of the two isotopes, since there are no branching points involved and the isotopes are far from neutron magic numbers. The $s$-process flux between the two Pd isotopes reaches the equilibrium defined by $N \sigma \simeq constant$, where $N$ is the abundance and $\sigma$ the neutron-capture cross section. The $^{135}$Cs/$^{133}$Cs ratio is higher in Monash by about $50\%$. This difference is probably due to different nuclear input physics, as the decay rate of $^{134}$Cs is constant in the Monash models but has a temperature dependency in the FRUITY models. The $^{182}$Hf/$^{180}$Hf ratio instead is more than one order of magnitude lower in the FRUITY models than in Monash. This is due to the different $^{181}$Hf $\beta$-decay rate used in the models.

\begin{table}
\caption{Comparison between the yields (in notation where, e.g., 7.43e-09 represents $7.43 \times 10^{-09}$) from a 3 \msun, $Z=0.014$ AGB model of Monash and FRUITY for $^{107}$Pd, $^{135}$Cs, and $^{182}$Hf as well as their respective stable reference isotopes.}
\centering
\begin{tabular}{l c c c c}
    Isotope & Monash & FRUITY \\
        \hline \hline
    $^{107}$Pd & 7.43e-09 & 2.64e-09 \\
    $^{108}$Pd & 5.17e-08 & 2.02e-08 \\
    $^{133}$Cs & 1.75e-08 & 8.58e-09 \\
    $^{135}$Cs & 5.83e-09 & 2.01e-09 \\
    $^{180}$Hf & 1.35e-08 & 3.27e-09 \\
    $^{182}$Hf & 1.71e-09 & 7.27e-11 \\
    \hline
    Ratio & & & & \\
    \hline
    $^{107}$Pd/$^{108}$Pd & 1.44e-01 & 1.31e-01 \\
    $^{135}$Cs/$^{133}$Cs & 3.33e-01 & 2.34e-01 \\
    $^{182}$Hf/$^{180}$Hf & 1.27e-01 & 2.32e-02 \\
    \hline
\end{tabular}
\label{tab:yields}
\end{table}

\subsection{GCE calibration} \label{sec:models}

The calibration of our GCE simulation is performed as in Paper I. Using observations of the Galactic disk \citep{kub15}, three GCE setups are considered for each set of stellar yields to reproduce a high, low, and best fit value for the radioactive-to-stable abundance ratios in the ISM at $t_{\odot}$. For both sets of yields, these GCE setups represent three distinct simulations of the Milky Way's disk, each with their own chemical evolution history. Each model is paramaterized using current observations and their associated uncertainties of the SFR, gas inflow rate, supernovae rate, and total mass of gas in the Galaxy. All models reach solar metallicity ($Z=0.014$) at the time of the birth of the Sun. We do not look at individual elements and isotopes to calibrate our GCE setups as it is beyond the scope of the paper to explore isotopes other than $s$-only, since these have contributions from several astrophysical sources for which we would need to explore different yields options. In any case, we note that a similar \texttt{OMEGA+} GCE setup to those in this paper is used by \cite{jones19} wherein they find good agreement with the solar abundances for several isotopes from Si to Nb. The adopted values of $A_1$ and $A_2$; the star formation efficiency, $f_{\star}$; and the mass loading parameter, $\eta$, for each setup are shown in Table \ref{tab:param} (see Section \ref{sec:omega} for a discussion of these parameters). Below, we give a brief summary of the reasoning behind our choice of the values for each parameter though we refer to Paper I for a more detailed explanation.

\begin{table}
\centering
\caption{Adopted values of parameters in our GCE framework for our low, best, and high fit models.}
\begin{tabular} {cccc}

                           Quantity & Low & Best & High \\ 
                          \hline\hline
                          $A_1$ [ \msun\ yr$^{-1}]$ & 91 & 46 & 0.5 \\  
                          $A_2$ [ \msun\ yr$^{-1}]$ & 2.9 & 5.4 & 10.0 \\ 
                          $f_{\star}$ $[10^{-10}$yr$^{-1}]$ & 1.8 & 2.6 & 6.5 \\
                          $\eta$ & 0.45 & 0.50 & 0.45\\
                          \hline
                          
\end{tabular}
\label{tab:param}
\end{table}

To increase the amount of stable isotopes in the ISM at $t_{\odot}$, and therefore obtain the low value of the radioactive-to-stable ratio, the first infall normalization parameter $(A_1)$ is increased, whereas the second infall normalization parameter ($A_2$) is decreased. This increases the magnitude of the first gas infall episode and decreases the magnitude of the second, so that we reach the upper limit for the observed stellar mass in the Galaxy, whilst also reaching the lower limit for the observed inflow rate. By increasing the star formation process at earlier times, more stable isotopes are produced by $t_{\odot}$, however, the SLR abundances at $t_{\odot}$ remain mostly unchanged as they are more sensitive to the SFR rather than the total integrated star formation history \citep{cote19}. For this reason, the high radioactive-to-stable GCE setup has a small $A_1$ and large $A_2$. Increasing the star formation efficiency means that more stable isotopes are locked inside stars, thus decreasing the gas-to-star ratio. Therefore, the higher the star formation efficiency the higher the radioactive-to-stable abundance ratio at $t_{\odot}$. The value of $\eta$, which determines the magnitude of the Galactic outflows, is chosen to remove enough metals from the Galaxy in order to recover solar metallicity at $t_{\odot}$.

\begin{table}
\caption{Estimated percentage of $^{108}$Pd, $^{133}$Cs, and $^{180}$Hf from the \textit{s}- and \textit{r}-process in the ISM at $t_{\odot}$. The roman and the italics numbers are calculated using Monash and FRUITY yields, respectively. The \textit{s}-process contribution from AGB (main) and massive stars (weak) is shown separately.}
\begin{tabular} {c c c c c c}

                            \multirow{2}{*}{Isotope} & \multirow{2}{*}{GCE}  & \multicolumn{2}{c}{\textit{s}-process} & {\textit{r}-process} \\
                           &  & \multicolumn{1}{c}{main} & \multicolumn{1}{c}{weak} &  \\
                           \hline
                           \hline
                            
                         \multirow{3}{*}{$^{108}$Pd\footnote{Note that the $s$- and $r$-process may not be the only sources of Pd in the Galaxy (see Section \ref{sec:lepp})}} 
                         &                            High & $59,\textit{90}$ & $1$ & $40,\textit{09}$ \\
                         &                            Best & $44,\textit{60}$ & $1$ & $55,\textit{39}$ \\ 
                         &                            Low  & $36,\textit{46}$ & $1$ & $63,\textit{53}$ \\
                         \hline
                         \multirow{3}{*}{$^{133}$Cs} & High & $20,\textit{37}$ & $1$ & $79,\textit{62}$ \\
                         &                             Best & $16,\textit{25}$ & $1$ & $83,\textit{74}$ \\ 
                         &                             Low  & $14,\textit{21}$ & $\sim0$ & $86,\textit{79}$ \\
                         \hline
                         \multirow{3}{*}{$^{180}$Hf} & High & $84,\textit{84}$ & $1$ & $15,\textit{15}$ \\
                         &                             Best & $85,\textit{83}$ & $1$ & $14,\textit{16}$ \\ 
                         &                             Low  & $88,\textit{83}$ & $\sim0$ & $12,\textit{17}$ \\
                         \hline

\end{tabular}
\label{tab:residual}
\centering
\end{table}

\begin{figure*}
    \centering
    \includegraphics[width=.48\linewidth]{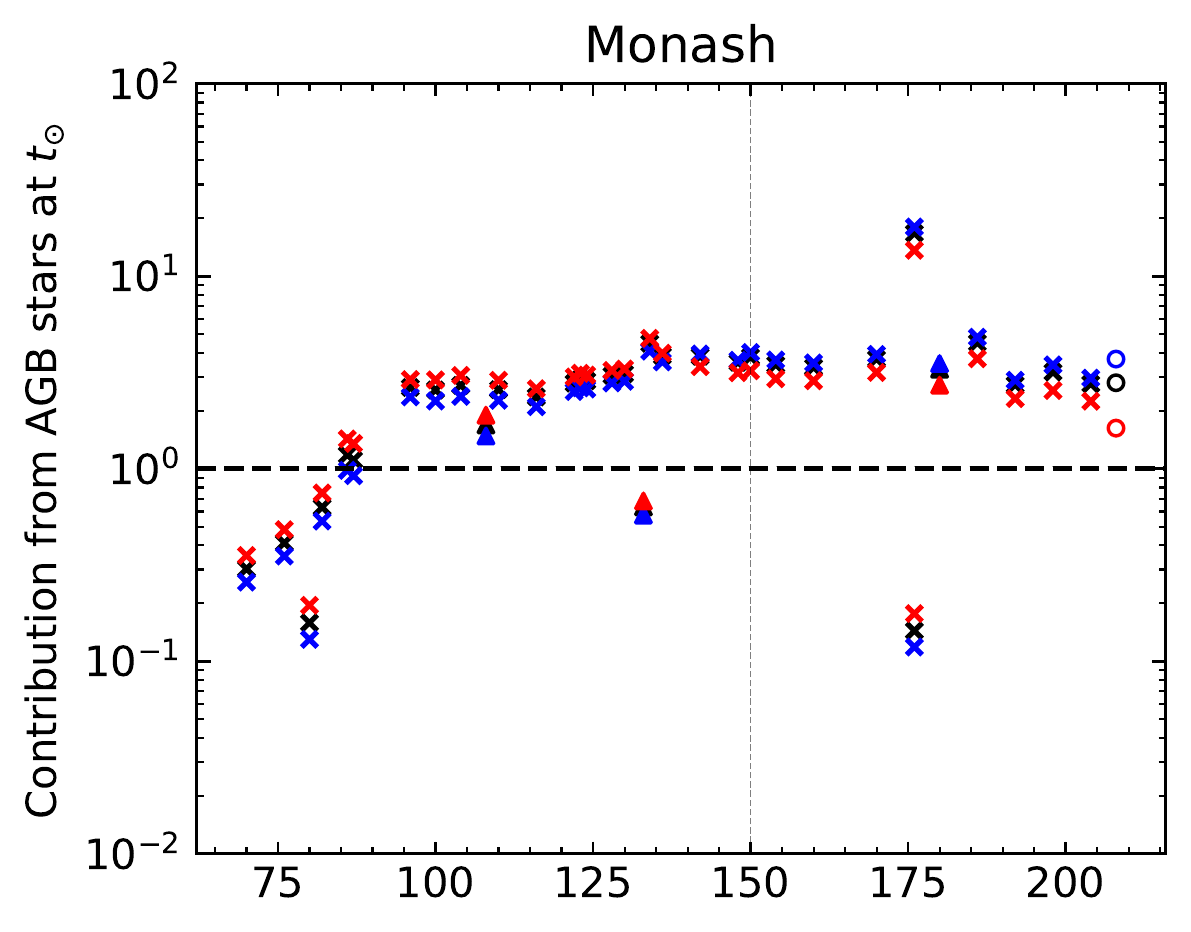} 
    \includegraphics[width=.46\linewidth]{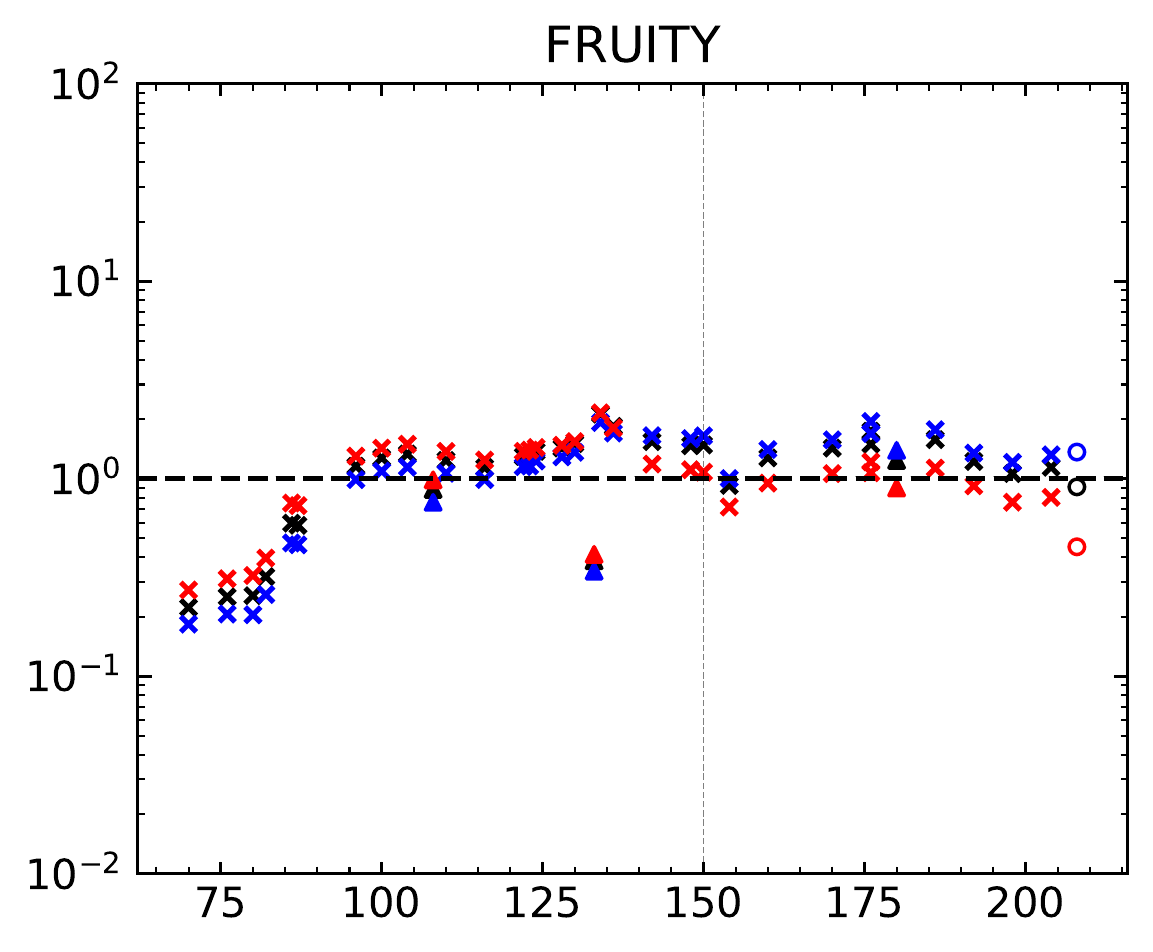} 
    \includegraphics[width=.48\linewidth]{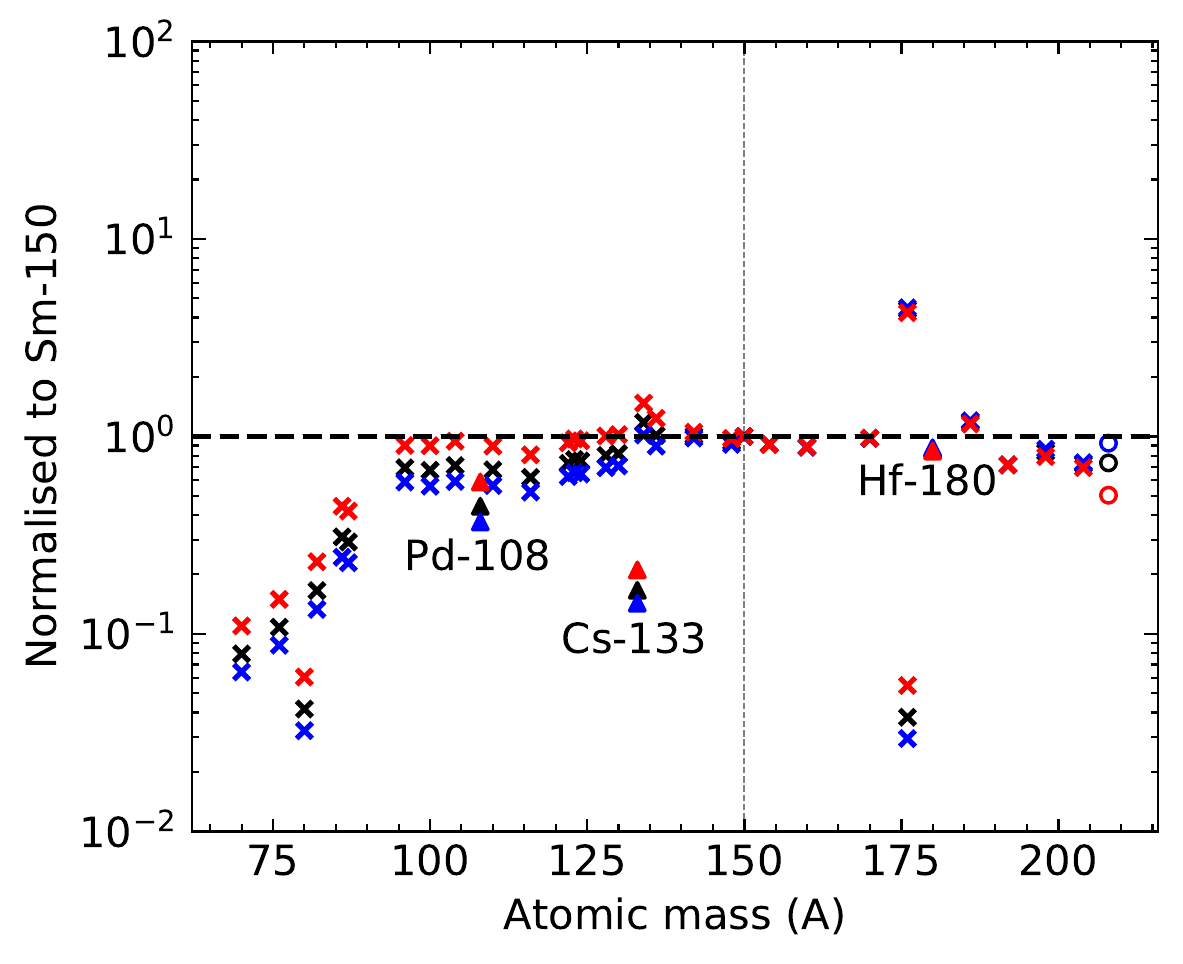} 
    \includegraphics[width=.46\linewidth]{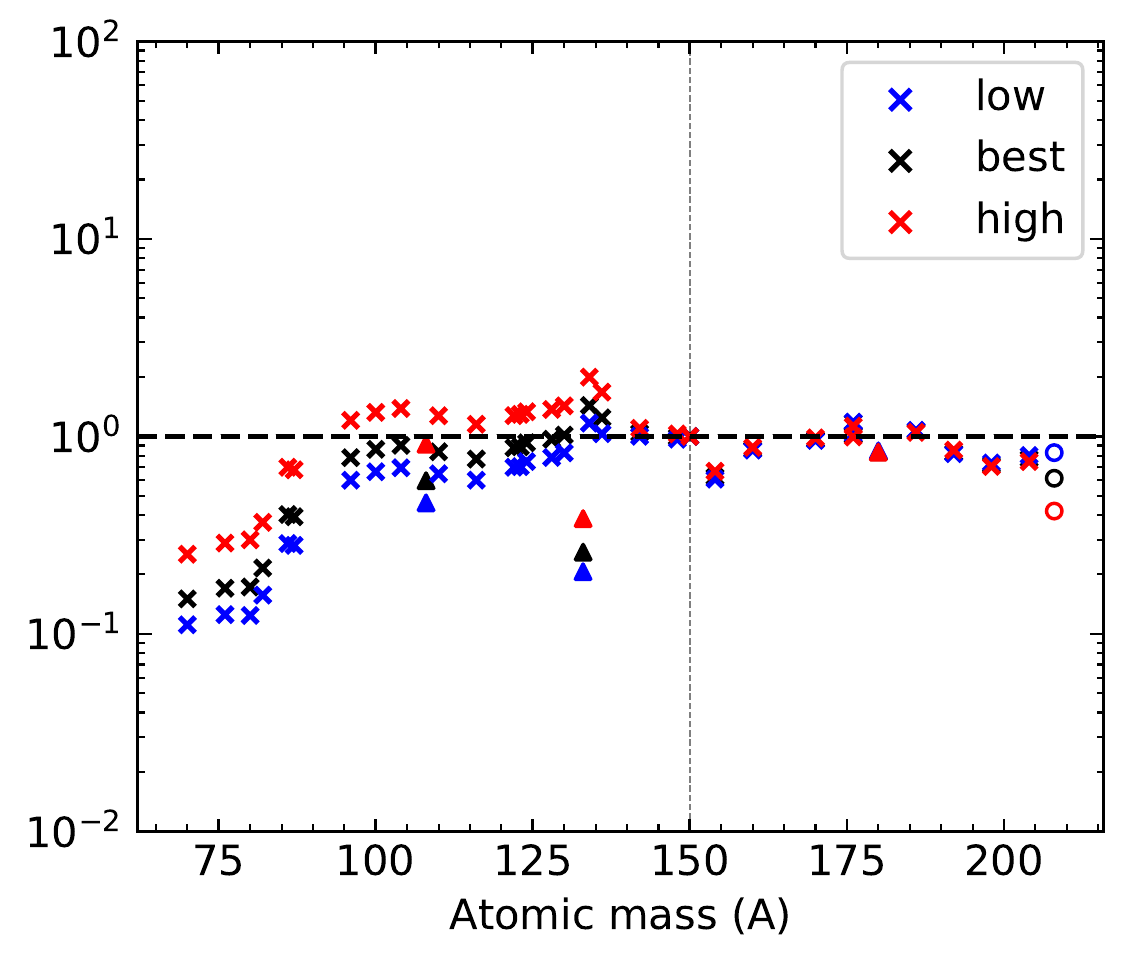} 
    \caption{GCE production factors due to AGB stars for the stable $s$-only isotopes (marked by crosses) and $^{208}$Pb (marked by a circle) in our low (blue), best (black), and high (red) Milky Way models (top panel) using the Monash and FRUITY yields, and then normalised to $^{150}$Sm (bottom panel). The stable reference isotopes ($^{108}$Pd, $^{133}$Cs, and $^{180}$Hf) are marked by a triangle. Note that in the Monash models, not all the \textit{s}-process branching points are properly implemented, which results in unrealistic predictions for some isotopes (e.g., $^{176}$Lu and $^{176}$Hf).}
    \label{fig:prod_factor}
\end{figure*}

To see how well the GCE setups reproduce the solar $s$-isotopic distribution at $t_{\odot}$, in the top panels of Figure \ref{fig:prod_factor} we show the isotopic distribution in the ISM at $t_{\odot}$ of the \textit{s}-only\footnote{We note that $^{80}$Kr and $^{86}$Sr in particular may attribute an appreciable fraction of their ISM abundance to the $p$-process \citep{travaglio:2015}} stable nuclei: $^{70}$Ge, $^{76}$Se, $^{80,82}$Kr, $^{86,87}$Sr, $^{96}$Mo, $^{100}$Ru, $^{104}$Pd, $^{110}$Cd, $^{116}$Sn, $^{122,123,124}$Te $^{128,130}$Xe, $^{134,136}$Ba, $^{142}$Nd, $^{148,150}$Sm, $^{154}$Gd, $^{160}$Dy, $^{170}$Yb, $^{176}$Lu, $^{176}$Hf, $^{186,187}$Os, $^{192}$Pt, $^{198}$Hg, $^{204}$Pb, as well as $^{208}$Pb (not $s$-only) and the stable isotopes of interest here that have both an $s$- and $r$-process origin, $^{108}$Pd, $^{133}$Cs, and $^{180}$Hf. All isotopes in Figure \ref{fig:prod_factor} are plotted relative to their present day solar values \citep{lod09}. In general, the GCE setups using Monash yields overproduce the $s$-isotopes relative to solar, except for $A<90$ which are mainly produced by massive stars. The GCE setups that use the FRUITY AGB yields produce less $s$-isotopes in the ISM at $t_{\odot}$ than Monash and thus obtain a better match with solar. Figure \ref{fig:yields} confirms that the FRUITY AGB yields for $^{107,108}$Pd, $^{133,135}$Cs, and $^{180,182}$Hf are lower than their Monash counterparts, suggesting that the physical conditions in the Monash models lead to a more efficient $s$-process. Here we are not overly concerned with reproducing the absolute solar abundances for the $s$-isotopes, as we are instead more interested in relative abundance ratios. However, the stable reference isotopes $^{108}$Pd, $^{133}$Cs, and $^{180}$Hf also have an $r$-process contribution in the ISM which we did not include in our GCE framework. This needs to be evaluated and taken into account when calculating the isotopic ratios of Pd, Cs, and Hf. To determine the $r$-process residual, we have to derive what fraction of the respective ESS abundance of the stable reference isotopes has an $s$-process origin once the $s$-only distribution is normalised relative to solar. 

To normalise the distribution to the solar values we scale the abundances relative to that of the \textit{s}-only isotope $^{150}$Sm. The reason for this is twofold: first, due to the short half-lives of $^{149}$Nd $(T_{1/2}=1.73\text{ hours})$ and $^{149}$Pm $(T_{1/2}=53.08\text{ hours})$, the complete $s$-process flow proceeds through $^{150}$Sm, regardless of whether or not the nearby branching points at $^{148}$Pm and $^{147}$Nd are activated \citep[see e.g.][]{arlandini99, bis15}; second, like other rare earth elements, the solar abundance of Sm is measured to a high precision. 

We do not consider an $r$-process contribution for $^{107}$Pd, $^{135}$Cs, and $^{182}$Hf, because the last \textit{r}-process event has been inferred to have occurred at least 100 Myr before $t_{\odot}$ \citep{lugaro14, tsuj17, cote2021}. This means that the SLRs would have significantly decayed in the local ISM from that event by $t_{\odot}$. Using the GCE setups normalised to $^{150}$Sm, the \textit{r}-process residual $Y$ for a stable isotope produced both by the \textit{s}- and \textit{r}-process, is calculated as

\begin{equation}
    Y_{\text{residual}} = 1 - (Y_{\text{MS}}/Y_{\odot} + Y_{\text{AGB}}/Y_{\odot}),
\label{eq:res}
\end{equation}
where the subscripts MS and AGB denote the ISM contribution in solar masses from massive and AGB stars respectively, and the subscript $\odot$ the solar abundance. The $s$-process yields from massive stars are included in the core-collapse supernovae yields \citep{rit18} used as input for the GCE simulations. The yields from each type of stellar source are tracked independently in \texttt{OMEGA+}, which means we can see the relative contribution from each source to the total mass of an isotope in the ISM. The relative \textit{s}-process contribution from massive and AGB stars for the reference isotopes $^{108}$Pd, $^{133}$Cs, and $^{180}$Hf in each GCE model are shown in Table \ref{tab:residual}. 

The $r$-residuals of $^{180}$Hf and $^{133}$Cs are relatively unchanged for the two AGB data sets, with a $12\%-17\%$ $r$-process contribution in the case of $^{180}$Hf, and a $62-86\%$ contribution for $^{133}$Cs, which are consistent with those obtained by \citep{pra20}: $19\%$ and $84\%$, respectively. However, the $r$-residual of $^{108}$Pd is significantly affected by changes in the choice of GCE parameters, particular when using the FRUITY AGB yields. This can be explained by considering the fact that since $^{108}$Pd is a first peak $s$-isotope, its production at higher metallicities is favoured relative to $^{133}$Cs and $^{180}$Hf which require a higher neutron-to-seed ratio (i.e., a higher $^{13}$C/$^{56}$Fe ratio). By changing the GCE setup, we change the SFR and the evolution of metallicity in the Milky Way model, which consequently changes the relative weight of each metallicity (in the AGB yields) on the overall Solar System enrichment. In the high GCE setup the gas metallicity evolves the fastest (see Figures 6 and 7 in Paper I), which puts the most weight on high-Z AGB models and consequently the production of the first peak $s$-isotopes. \cite{cristallo2015b} also observe an increase in the production of light $s$-isotopes, relative to heavier ones, in their GCE models that use an increased SFR.

Furthermore, when determining the $r$-residual of $^{108}$Pd we must be also be mindful of the well known problem that, when considering contributions from only the $s$- and $r$-process, the abundances of the elements around the first $s$-process peak (Sr, Y, Zr, close to Pd) are underproduced compared to solar \citep{trav04}. For this reason, \cite{trav04} invoke an additional process, the \textit{lighter element primary process} (LEPP), in order to account for the missing fraction of the abundances of these isotopes in the ISM; the existence of a LEPP is still the subject of much debate (see Section \ref{sec:dis}). When using the FRUITY yields \cite{cristallo2015b} did not find this problem and this result is confirmed by our results (see the best and high GCE setups in the bottom right panel of Figure \ref{fig:prod_factor}). However, a problem we encounter is that the setups that match the first $s$-process peak (i.e., the best and the high FRUITY setups and the high Monash setup) under-produce the third $s$-process peak, i.e. $^{208}$Pb, by a factor of two \citep[see also][]{travaglio01}.

The low FRUITY GCE setup and the low and best Monash setups instead produce sub-solar abundances (i.e., below the dashed line in the bottom panel of Figure \ref{fig:prod_factor}) for Sr, Y, Zr, while $^{208}$Pb is closer to solar. Therefore, we would still need to invoke an additional contribution of the first $s$-process peak (e.g., a LEPP) in these models. We readdress this problem in Section \ref{sec:lepp} using for the first time the isotopic ratio $^{107}$Pd/$^{182}$Hf.

\section{Results} \label{sec:results}

In this section, we present the evolution of the abundance ratios related to the SLRs of interest, assuming an homogeneous distribution of isotopes in the Galaxy (see Section \ref{sec:time}.) Then, we derive the isolation times from each of these abundance ratios (see Section \ref{sec:iso}). We apply the statistical analysis in Papers II and III to calculate the uncertainties on the isolation time, $T_{\text{iso}}$, due to the effects of heterogeneities on SLR abundances in the ISM assuming that $\tau/\gamma>2$ (Regime I). In this Regime, the average length of time between the formation of enrichment progenitor $\gamma$ is short enough that the SLR does not have time to completely decay in the ISM, so a GCE description is valid. In Section \ref{sec:lepp}, we look at the predicted $^{107}$Pd/$^{182}$Hf abundance ratio and use it to estimate the amount of $^{107}$Pd in our GCE setups with a non $s$- or $r$-process origin. In Section \ref{sec:LE} we consider instead the case where $\tau/\gamma\lesssim0.3$ and then derive times from the last event and potential constraints on the nucleosynthesis in this last event.

\subsection{Time evolution of the SLR ratios} \label{sec:time}

\begin{figure*}
    \includegraphics[width=.49\linewidth]{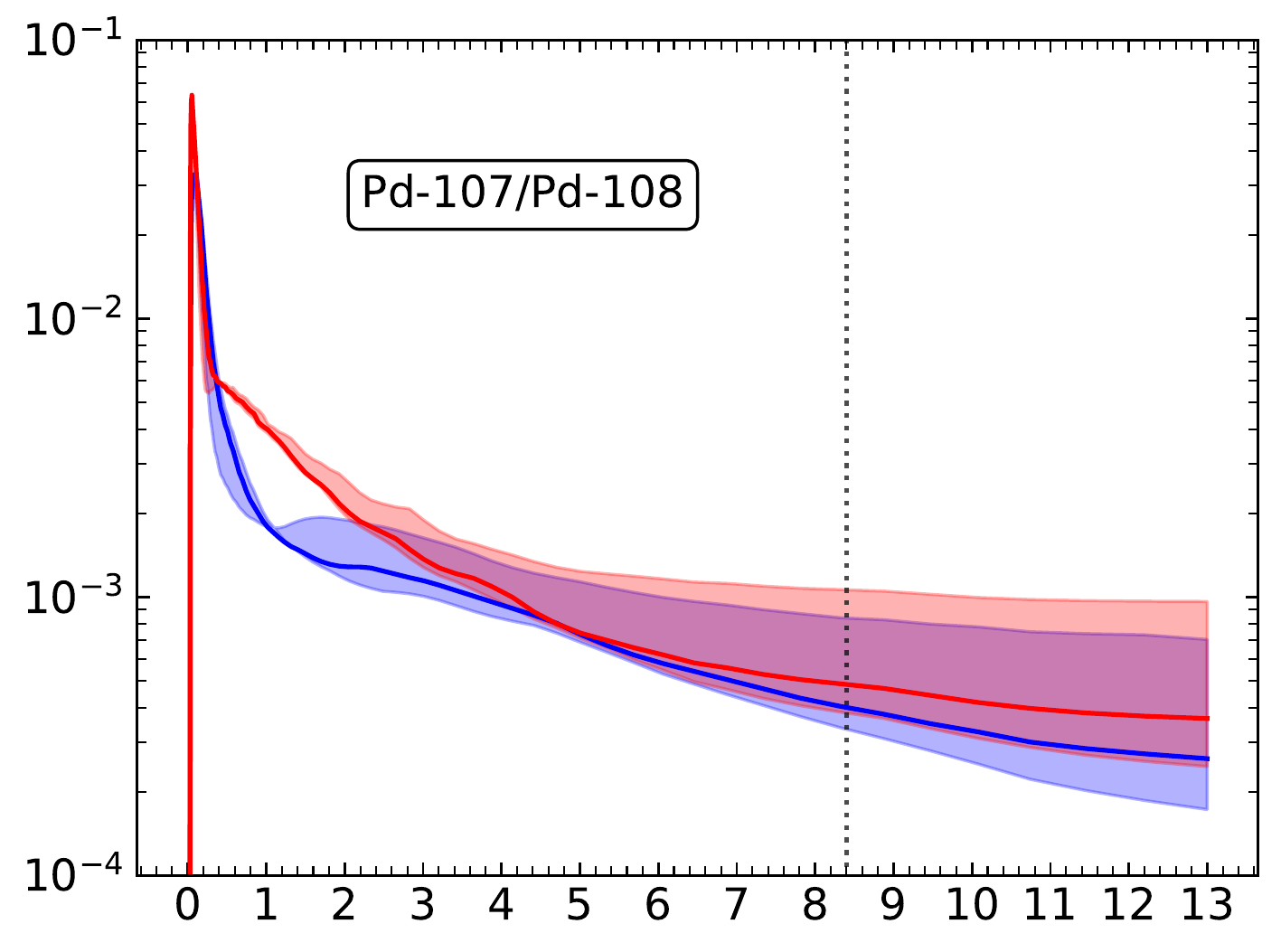}
    \includegraphics[width=.49\linewidth]{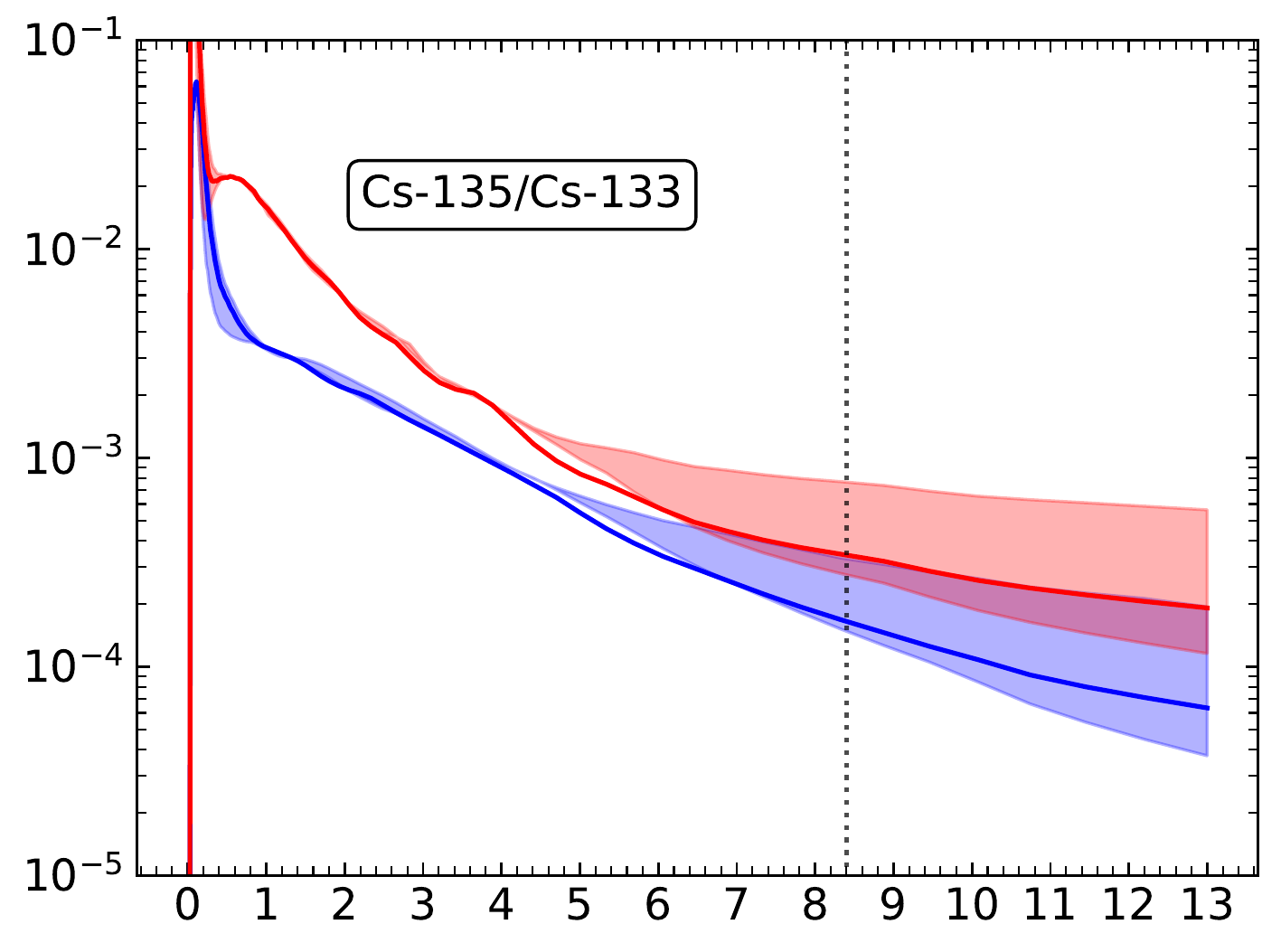} 
    
    \includegraphics[width=.49\linewidth]{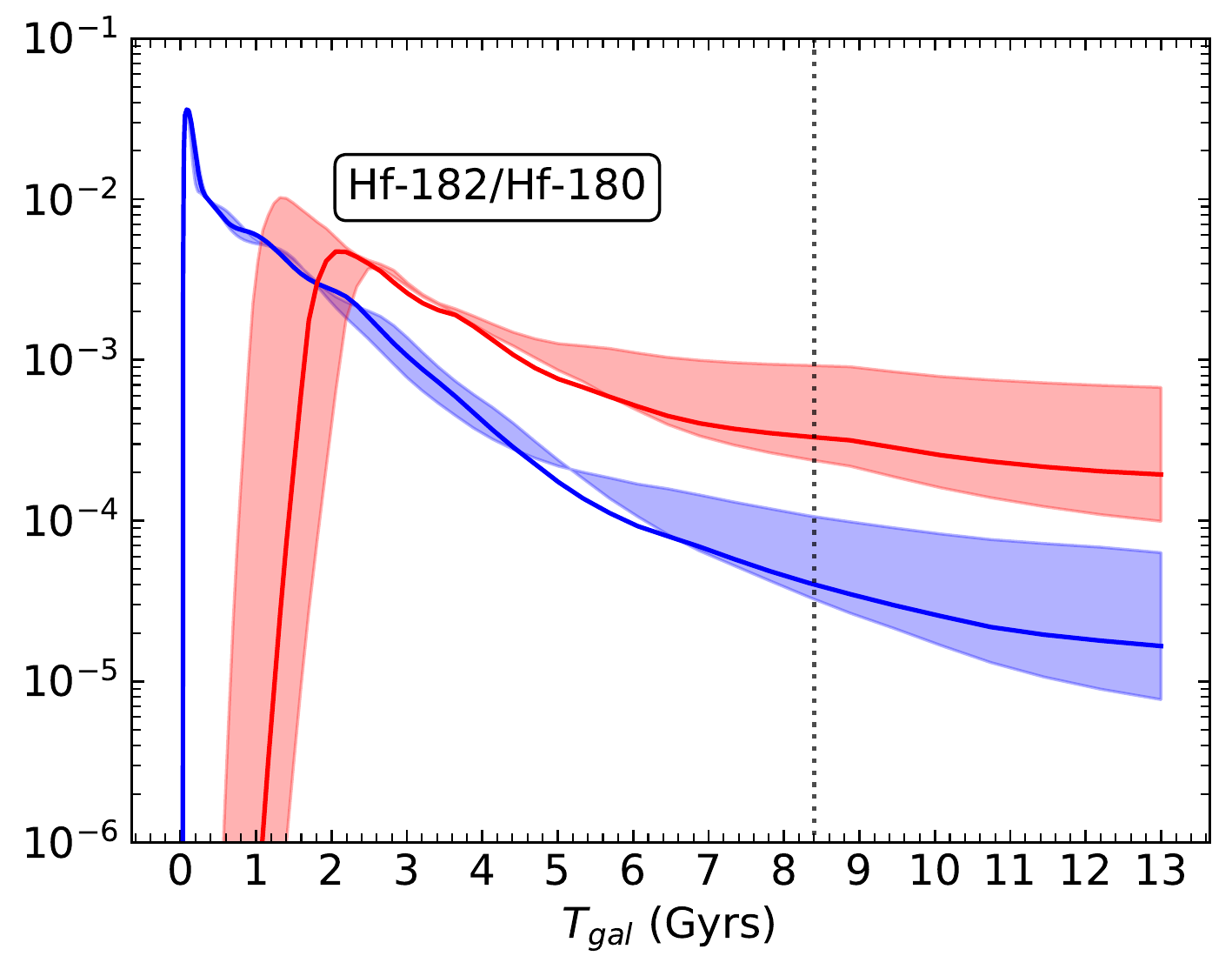}
    \includegraphics[width=.49\linewidth]{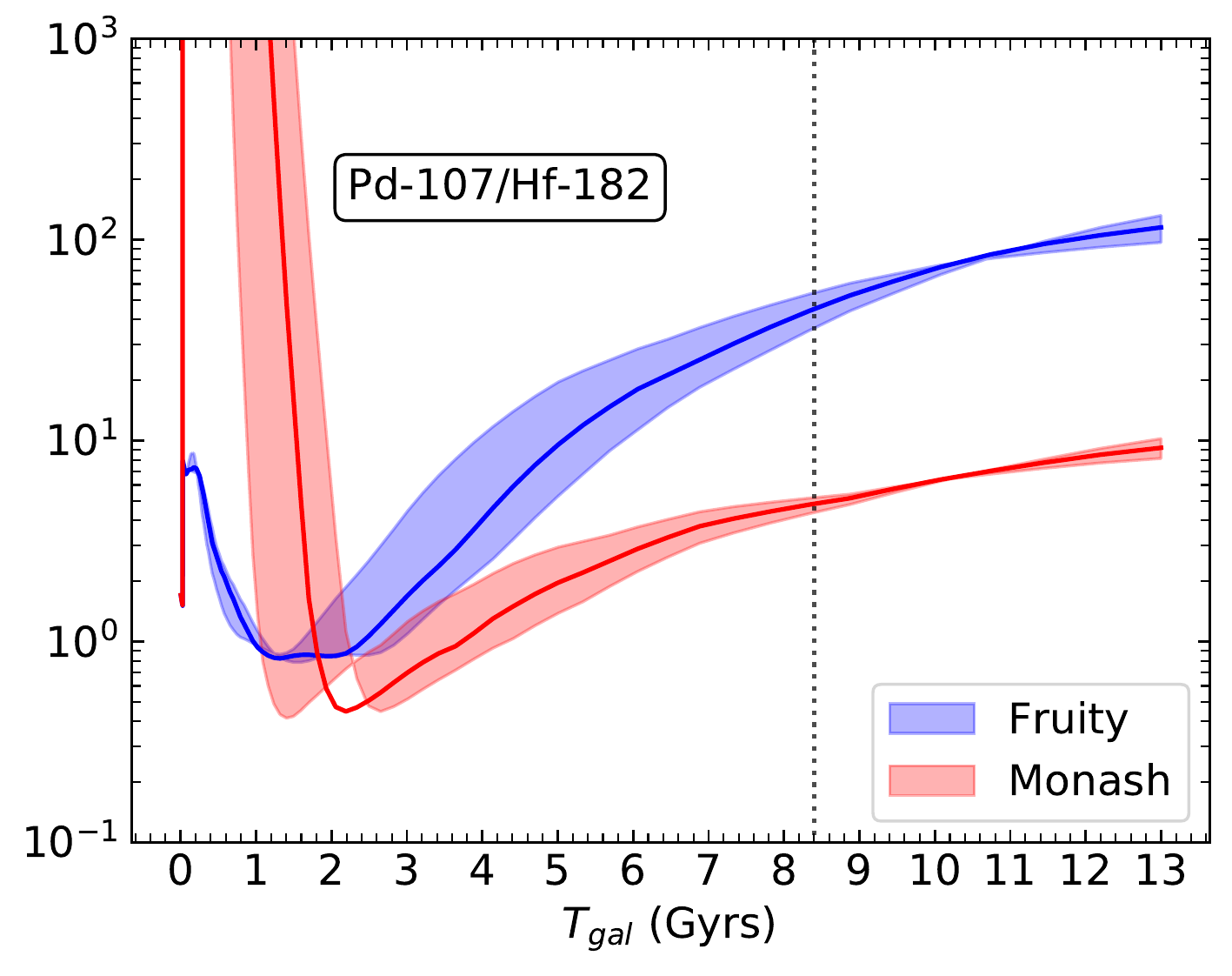} 
    \caption{Evolution of the $^{107}$Pd/$^{108}$Pd, $^{135}$Cs/$^{133}$Cs, $^{182}$Hf/$^{180}$Hf, and $^{107}$Pd/$^{182}$Hf abundance ratios in the ISM as a function of Galactic time using Monash (red) and FRUITY (blue) ABG stellar yields. For each set of yields, the best (solid line), and high and low (extremities of the shaded region) GCE setups are plotted. The vertical dashed line is at $t_{\odot}=8.4$ Gyr. The ratios do not include the $r$-process residual for the stable isotope since the residual is calculated only at $t_{\odot}$.}
    \label{fig:ratio}
\end{figure*}

In Figure \ref{fig:ratio} the time evolution of three radioactive-to-stable abundance ratios and the ratio of two SLRs, $^{107}$Pd/$^{182}$Hf, are plotted for the high, low, and best fit GCE setups. These ratios take into consideration only the $s$-process contributions in the ISM. The dashed vertical line at $8.4$ Gyr indicates $t_{\odot}$, corresponding to the time when $Z=Z_{\odot}=0.014$. Prior to calculating $T_{\text{iso}}$ from each ratio, we added the respective \textit{r}-process residual from Table \ref{tab:residual} to the reference isotope at $t_{\odot}$\footnote{The residual is not added in Figure \ref{fig:ratio} since it applies only at $t_{\odot}$ specifically, not at all Galactic times.}. Recall that the GCE setups, represented by the coloured bands in the figure, represent unique solutions for the chemical evolution of the Milky Way. Each solution has its own star formation history and Galactic inflows and outflows, such that the width of each colour band at $t_{\odot}=8.4$ Gyr represents the associated GCE uncertainties on the radioactive-to-stable abundance ratios in the ESS for each set of yields. However, they do not necessarily represent the upper and lower limits on the ISM ratio at earlier or later Galactic times. Since our low and high GCE setups are calibrated to minimise and maximise the ISM ratio at $t_{\odot}$, respectively, it would be incorrect, for example, to compare the abundance ratio from the best fit model of FRUITY with the abundance ratio from the low fit model of Monash, as they represent different Galaxies entirely. 

The temporal evolution of the $^{107}$Pd/$^{108}$Pd, $^{135}$Cs/$^{133}$Cs, and $^{182}$Hf/$^{180}$Hf ratios in Figure \ref{fig:ratio} are typical for SLRs in the ISM. The abundance ratio peaks at early galactic times, as the gas inflow rate and subsequently also the SFR peak then. When the SFR drops, the SLR reaches a steady-state abundance in the ISM when its stellar production rate is balanced by its decay rate. The abundance of the stable isotope instead continues to rise and therefore the radioactive-to-stable abundance ratios decrease with time.

The evolution of the $^{107}$Pd/$^{108}$Pd abundance ratio is very similar in the Monash and FRUITY GCE setups. This result reflects the similarity in the mass and metallicity dependent yields for the two Pd isotopes from the two sets of stellar models shown in Figure \ref{fig:yields}. The higher $^{135}$Cs/$^{133}$Cs abundance ratio in the Monash setups is because the Monash stellar models produce more $^{135}$Cs than FRUITY at the same initial metallicity (see Figure \ref{fig:yields}). However, this difference between the $^{135}\text{Cs}$/$^{133}\text{Cs}$ abundance ratios of the two data sets is reduced when the $r$-process residual of $^{133}$Cs is added (as listed in Table \ref{tab:residual}), because the $r$-process residuals in each of the Monash GCE setups are higher than those for FRUITY. 

Regarding the evolution of the $^{182}$Hf/$^{180}$Hf abundance ratio, the production of $^{182}$Hf is delayed for Monash with respect to FRUITY because this isotope is not included in the lowest metallicity Monash models. However, this has negligible consequence for the abundance ratio at $t_{\odot}$, since only the $^{182}$Hf produced by the most recent events is potentially inherited by the ESS. The Monash abundance ratio at $t_{\odot}$ is almost an order of magnitude higher than FRUITY due to the significantly longer half-life (see Section \ref{sec:yields}) used for $^{181}$Hf in the Monash models, which leads to a much larger production of $^{182}$Hf.

In Figure \ref{fig:ratio} we also show the abundance evolution of $^{107}$Pd relative to $^{182}$Hf; the former is representative of the first \textit{s}-process peak, whilst the latter of the second peak. Interestingly, although $^{107}$Pd has a shorter half-life than $^{182}$Hf, the ratio increases with time because $^{107}$Pd is produced more than $^{182}$Hf by higher metallicity ($Z\sim0.014$) AGB stars in both sets of models. This result can be directly attributed to the fact that, unlike the iron peak $s$-process seeds (notably $^{56}$Fe), $^{13}$C is a primary neutron source, since it is produced inside the star starting from the initial H and He \citep[see e.g.][]{gal98}. Therefore, AGB stars with lower initial metallicities have a higher neutron-to-seed ratio (i.e., $^{13}$C/$^{56}$Fe) and, consequently, produce more effectively nuclei beyond the first \textit{s}-peak. Observations of Ba stars, the binary companion of AGB stars, confirm this trend \citep{cseh18}. The very high $^{107}$Pd/$^{182}$Hf ratio in the FRUITY setups compared to Monash setups originates again from the higher $^{181}$Hf $\beta$-decay rate used to calculate these models. 

\subsection{Derivation of the isolation times} \label{sec:iso}

For each abundance ratio plotted in Figure \ref{fig:ratio}, $T_{\text{iso}}$ is the time taken for the predicted ratio at $t_{\odot}$ to decay to the observed ESS value, assuming the ESS is not exposed to further stellar enrichment events. This is a reasonable assumption as the evolutionary stages of low-mass stars, the main $s$-process sources in the Galaxy, prior to the AGB phase have a long duration ($\sim 1$ Gyr), however, star-forming regions typically live at most for a few tens of Myr. Therefore these regions would have dissolved by the time a low-mass star reaches the AGB. Also the probability of an AGB star encountering a star-forming region is extremely low \citep{kas94}. We reiterate that the values for $T_{\text{iso}}$ calculated using the low and high GCE setups represent the associated GCE uncertainty from observational constraints of the Milky Way disc. 

Since \texttt{OMEGA} assumes a smooth and continuous enrichment process, we must determine the associated error on $T_{\text{iso}}$ that arises when we instead consider that stellar additions are discrete in time.

\begin{table*}
\caption{Isolation times $T_{\text{iso}}$ calculated from the abundance ratios at $t_{\odot}$ from Figure \ref{fig:ratio} assuming that they fall into Regime I ($\tau/\gamma\gtrsim2$). The ISM ratios are taken from Figure \ref{fig:ratio} with an additional $r$-process residual added to the stable isotopes according to Table \ref{tab:residual}. The uncertainty factors for the isolation times are calculated following the statistical analysis presented in Paper II. We do not calculate $T_{\text{iso}}$ using the $^{107}$Pd/$^{182}$Hf ratio, since a fraction of the Pd isotopes in the ESS may have a non $s$- or $r$-process origin (see Section \ref{sec:lepp}), however, Hf does not. All $\tau$, $\gamma$, and $T_{\text{iso}}$ are given in Myr. The value of $\tau$ for $^{107}$Pd/$^{182}$Hf represents the equivalent $\tau$ of the two SLRs, where $\tau_{\text{eq}}=\tau_1\tau_2/(\tau_1-\tau_2)$. The uncertainty factors for this ratio are taken from Paper III.}
\begin{tabular}{c c c c c c}
 & & $^{107}$Pd/$^{108}$Pd & $^{135}$Cs/$^{133}$Cs & $^{182}$Hf/$^{180}$Hf & $^{107}$Pd/$^{182}$Hf \\
 \\[-1em]
 \hline\hline
ESS Ratio & & $6.6\times10^{-5}$ & $<2.8\times10^{-6}$\footnote{From recent measurements of Ba isotopic abundances in chondrites it was inferred that an upper limit for this ratio is $4.6\times10^{-4}$ \citep{sak20}. This ratio is higher than that predicted in each of our GCE setups, so the only possible solution is $T_{\text{iso}}>0$, which does not provide any constraint.} & $1.02\times10^{-4}$ & 4.25 \\
$\tau$    & & 9.4 & 3.3 & 12.8 & 35.4 \\
$\gamma$ (adopted)  & & 3.16 & 1.00 & 3.16 & 10.00 \\
$\tau/\gamma$ & & 2.97 & 3.30 & 4.05 & 3.54 \\
Uncertainty Factors & & $0.61, 1.39$ & $0.63, 1.45$ & $0.61, 1.39$ & $0.73, 1.17$ \\
\hline
\\[-1em]
\multicolumn{2}{c}{GCE Model} & \multicolumn{4}{c}{ISM Ratio}\\
\\[-1em]
\hline
 Low & Monash & $1.7\times10^{-4}$ & $5.2\times10^{-5}$ & $2.2\times10^{-4}$ & 4.37 \\
          & FRUITY & $1.8\times10^{-4}$ & $4.0\times10^{-5}$ & $-$ & $-$ \\
          Best & Monash & $2.5\times10^{-4}$ & $7.9\times10^{-5}$ & $3.1\times10^{-4}$ & 4.83 \\
          & FRUITY & $2.7\times10^{-4}$ & $5.4\times10^{-5}$ & $-$ & $-$ \\
          High & Monash & $7.3\times10^{-4}$ & $2.2\times10^{-4}$ & $8.6\times10^{-4}$ & 5.19 \\
          & FRUITY & $7.7\times10^{-4}$ & $1.5\times10^{-4}$ & $-$ & $-$ \\
\hline
\\[-1em]
& & \multicolumn{4}{c}{Isolation Time}\\
\\[-1em]
\hline
 Low & Monash & $9^{+3}_{-5}$ & $>10^{+1}_{-1}$ & $10^{+4}_{-6}$ &  $-$ \\
          & FRUITY & $9^{+3}_{-5}$ & $>9^{+1}_{-1}$ & $-$ & $-$ \\
          Best & Monash & $13^{+3}_{-5}$ & $>11^{+1}_{-1}$ & $14^{+4}_{-6}$ & $-$ \\
          & FRUITY & $13^{+3}_{-5}$ & $>10^{+1}_{-1}$ & $-$ & $-$ \\
          High & Monash & $23^{+3}_{-5}$ & $>14^{+1}_{-1}$ & $27^{+4}_{-6}$ & $-$ \\
          & FRUITY & $23^{+3}_{-5}$ & $>13^{+1}_{-1}$ & $-$ & $-$ \\
\hline
\end{tabular}
\label{tab:iso}
\end{table*}

Table \ref{tab:iso} shows the values of $T_{\text{iso}}$ calculated using the ratios from Figure \ref{fig:ratio} along with the error analysis of Paper II. Paper II determined that it is possible to define a statistical distribution for the evolution of an SLR if it falls in the $\tau/\gamma\gtrsim2$ Regime (henceforth Regime I from Paper II). For an SLR in Regime I, the time between the formation of two progenitors $\gamma$ is short enough to ensure that its minimum abundance at $t_{\odot}$ is always greater than zero. In Paper II, the SLR ISM abundance uncertainty was calculated with a Monte Carlo method as a function of $\gamma$ and $\tau/\gamma$ for three box delay-time-distribution (DTD) functions, each with uniform probability as defined in Paper II. We choose the DTD function with the longest delay time, which applies in the case where the average time interval between the formation of a progenitor and subsequent ejection of material is approximately $5$ Gyr. Since the parameter $\gamma$ is poorly understood, we take the largest value of $\gamma$ for each SLR, from the six available values in Paper II, whilst still remaining in Regime I (i.e. $\tau/\gamma\gtrsim2$). This approach gives us the most conservative errors from the Monte Carlo Spread. Following this approach, we adopt a value of $\gamma=3.16$ for $^{107}$Pd and $^{182}$Hf from Table 4 of Paper II. $^{135}$Cs, instead, has a much lower $\tau=3.3$, therefore, the maximum value that we can choose is $\gamma=1$ in order to fall into Regime I. In principle, $\gamma$ should be consistent for all three SLRs given that they are ejected by the same stellar enrichment event, however, in this work the choice of $\gamma$ is only relevant for determining the $68\%$ confidence level.

From Table \ref{tab:iso}, the range of $T_{\text{iso}}$ derived with the Monash \textit{(FRUITY)} GCE setups are; low: $9-12$ \textit{(8}$-$\textit{12)} Myr, best: $10-16$ \textit{(9}$-$\textit{16)} Myr, high: $18-26$ \textit{(18}$-$\textit{26)} Myr. In each GCE setup all $T_{\text{iso}}$, when available, overlap within $1\sigma$ uncertainty, regardless of the adopted yields and isotopic ratios. The values from the FRUITY GCE setups are less well constrained since we cannot use the $^{182}$Hf/$^{180}$Hf ratio in this case. The $^{107}$Pd/$^{108}$Pd ratio in the ISM at $t_{\odot}$ shows remarkable agreement for Monash and FRUITY in the respective low, best, and high GCE setups. This is due to the fact  this ratio depends mostly only on the ratio of the cross sections. Furthermore, the nuclear uncertainties for $^{107}$Pd/$^{108}$Pd are very small. Monte Carlo sampling of a log-normal Gaussian following the methodology of \cite{lon10}, gives a $1\sigma$ and $2\sigma$ uncertainty for the branching ratio of $12.4\%$ and $26.8\%$, respectively; these are insignificant compared to the GCE uncertainties considered in this paper (i.e. for both sets of yields there is more than a factor of 2 difference in $T_{\text{iso}}$ derived from the low and high GCE setups).

A conservative estimate for the nuclear uncertainty associated with the $^{182}$Hf/$^{180}$Hf branching point can be estimated by considering the reaction rates that can maximise and minimise this ratio. The maximum $^{182}$Hf/$^{180}$Hf ratio can be obtained by taking the upper limits of the $^{180}$Hf$(n,\gamma)$ and $^{181}$Hf$(n,\gamma)$ reaction rates, and the lower limits of the $^{181}$Hf $\beta$-decay and $^{182}$Hf$(n,\gamma)$ rates, as this would maximise the $^{182}$Hf and minimise the $^{180}$Hf yields. Likewise, reversing the limits for the above reaction rates returns the minimum $^{182}$Hf/$^{180}$Hf ratio. We tested the effects of changing the reaction rates to the above mentioned limits using the uncertainties reported by \cite{rau02} for the n,g rates, and by \cite{lugaro14} for the decay rate of $^{181}$Hf (mostly coming from the presence of a state at 45 keV) using a Monash 3 \msun, $Z=0.014$ AGB model. We found that the $^{182}$Hf yield could be a factor of two lower than the presently adopted value. If we consider this most extreme scenario by applying a factor of two reduction to all the $^{182}$Hf yields, the $^{182}$Hf/$^{180}$Hf and $^{107}$Pd/$^{182}$Hf ratios in the ISM at $t_{\odot}$ will be a factor of two lower and a factor of two higher, respectively. Given that $T_{\text{iso}}=\tau (\ln{r_{\odot}}-\ln{r_{\text{ESS}}})$ - where $\tau$ is the mean-life of the SLR, $r_{\odot}$ is the radioactive-to-stable abundance ratio in the ISM at $t_{\odot}$, and $r_{\text{ESS}}$ is the ESS ratio - then a reduction of factor two of the $^{182}$Hf/$^{180}$Hf ratio in the ISM results in an $T_{\text{iso}}$ derived from this ratio that is $\approx9$ Myr shorter than the respective times given in Table \ref{tab:iso}. If this were the case, no value of $T_{\text{iso}}$ exists in the low and best Monash GCE setups that is consistent when derived using all three radioactive-to-stable ratios. However, for the high Monash GCE setup, $T_{\text{iso}}$ would still be self-consistent in the range $18-22$ Myr. Therefore, whilst highly unlikely, since a solution exists in the high GCE setup, such an extreme reduction in the $^{182}$Hf/$^{180}$Hf ratio due to nuclear uncertainties is not entirely ruled out by our results. Despite the seemingly large nuclear uncertainty relating to the $^{182}$Hf/$^{180}$Hf branching point, the method we apply above only gives us a maximum or minimum value of the ratio with no probability associated to it. To better understand the uncertainty associated with this branching ratio, it would be necessary to perform a statistical analysis to identify which combination of reaction rates are more or less likely. 

\subsection{The $^{107}$Pd/$^{182}$Hf ratio} \label{sec:lepp}

The $^{107}$Pd/$^{182}$Hf ratio is interesting as it is indicative of the relative abundance ratio of isotopes prior to and beyond the second $s$-process peak at the neutron magic number $N=82$. Several previous GCE studies that considered contributions from both \textit{s}- and \textit{r}-process sources appear to underproduce the \textit{s}-process elements with $90<A<130$ relative to solar abundances \citep{trav99, trav04, bis2014, bis17}. Therefore, we might not be able to reliably use the $^{107}$Pd/$^{182}$Hf ratio to determine $T_{\text{iso}}$ since $^{107}$Pd might be underproduced in our GCE framework. On the other hand, the GCE studies by \cite{pra18, pra20} find that the weak $s$-process in rotating massive stars provides a sufficient contribution to the $90<A<130$ $s$-isotopes, such that a LEPP is not needed to reproduce their solar abundances. $^{107}$Pd and $^{182}$Hf can provide fresh information on this issue because they only sample production in stars of around solar metallicity, as any abundance produced in the early Galaxy in low-metallicity objects would have decayed by the time of the formation of the Sun. We can quantitatively estimate to what extent (if at all) $^{107}$Pd is underproduced in our GCE framework at $t_{\odot}$ by determining the $^{107}$Pd/$^{182}$Hf ratio in the ISM at $t_{\odot}$ that is necessary to obtain a $T_{\text{iso}}$ consistent with those calculated using the $^{107}$Pd/$^{108}$Pd and $^{182}$Hf/$^{180}$Hf ratios in Section \ref{sec:iso}. This method can provide us an independent constraint on the production of the first $s$-process peak in the Galaxy at the time of the formation of the Sun. Since we are using the ESS ratio of two SLRs, any ``missing" component of $^{107}$Pd cannot be from the $r$-process, as the last $r$-process event was over $100$ Myr before $t_{\odot}$ (see Section \ref{sec:models}). Furthermore, in order for the $^{107}$Pd/$^{108}$Pd ratio to remain unchanged, any deficit of $^{107}$Pd in our GCE setups corresponds to an identical deficit of $^{108}$Pd. 

We consider only the Monash GCE setups here as the FRUITY AGB models do not produce enough $^{182}$Hf. In order to get the $^{107}$Pd/$^{182}\text{Hf}$ ratio in the ISM at $t_{\odot}$ that includes contributions from all production channels, henceforth $r_{\text{ISM}}$, we decay the ESS $^{107}$Pd/$^{182}\text{Hf}$ ratio backwards in time (i.e., we reverse the radioactive decay process) by $T_{\text{iso}}$. The ESS $^{107}$Pd/$^{182}\text{Hf}$ ratio inferred from meteoritic analysis is taken here to be 4.25 \citep{lugaro18}, and we use upper and lower limits for $T_{\text{iso}}$ that are derived using the $^{107}$Pd/$^{108}$Pd and $^{182}$Hf/$^{180}$Hf ratios in Table \ref{tab:iso}. It follows, that for each of the low, best, and high Monash GCE setups the amount of $^{107}$Pd missing in the ESS, $f_{\text{Pd}}$, is

\begin{equation}
f_{\text{Pd}}=\frac{r_{\rm{ISM}}-r_{\rm{GCE}}}{r_{\rm{GCE}}}, 
\end{equation}
where the denominator $r_{\rm{GCE}}$ is the respective $^{107}$Pd/$^{182}\text{Hf}$ ratio at $t_{\odot}$ predicted by the GCE framework given in Table $\ref{tab:iso}$.    

\begin{table}
\caption{Upper and lower limits for the isolation time, $T_{\text{iso}}$; the $^{107}$Pd/$^{182}$Hf ratio in the ISM at $t_{\odot}$ obtained by decaying back the ESS ratio ($4.25$) by $T_{\text{iso}}$, $r_{\text{ISM}}$; the $^{107}$Pd deficit in the ISM at $t_{\odot}$, $f_{\text{Pd}}$; and the $^{107}$Pd/$^{182}$Hf production ratio, $P$ derived using the results in Table \ref{tab:iso} (see text for details).}
\centering
\begin{tabular}{c c c c c}
    & $T_{\text{iso}}$ (Myr) & $r_{\text{ISM}}$ & $f_{\text{Pd}}$ & $P$ \\
    \hline
    \hline
    Low & $4-12$ & $4.77-6.00$ & $9-38\%$ & $5.55-11.19$ \\
    Best & $8-16$ & $5.34-6.73$ & $11-39\%$ & $6.21-12.55$ \\
    High & $21-26$ & $7.77-8.98$ & $50-73\%$ & $9.04-16.75$ \\
    \hline
\end{tabular}
\label{tab:lepp}
\end{table}

Table \ref{tab:lepp} shows the upper and lower limits for $T_{\text{iso}}$, the decayed back $r_{\text{ISM}}$, and the estimated $^{107}$Pd deficit in each of the low, best, and high Monash GCE setups. In the final column, we also calculate the upper and lower limits on the production ratio for $^{107}$Pd/$^{182}$Hf,

\begin{equation}
     P= r_{\text{ISM}} \frac{\tau_{\text{Hf}}}{\tau_{\text{Pd}}},
\end{equation}
where $\tau_{\text{Pd}}$ and $\tau_{\text{Hf}}$ are the mean-life of $^{107}$Pd and $^{182}$Hf, respectively. 

Assuming that we are in Regime I, that is $\tau/\gamma\gtrsim2$ (see Section \ref{sec:iso}), we can determine the minimum and maximum values of $P$ within a $68\%$ confidence level that are consistent with the upper and lower limits for $r_{\odot}$. By applying the relevant uncertainty factors from the fifth row of Table \ref{tab:iso}, the maximum and minimum values for $P$ are given by,

\begin{equation}
     P_{\text{max}}=\frac{r_{\text{ISM},\text{max}}}{0.73} \frac{\tau_{\text{Hf}}}{\tau_{\text{Pd}}},
\end{equation}
and
\begin{equation}
    P_{\text{min}}=\frac{r_{\text{ISM},\text{min}}}{1.17} \frac{\tau_{\text{Hf}}}{\tau_{\text{Pd}}},
\end{equation}
where we have used the subscripts \textit{min} and \textit{max} to denote the upper and lower limits on our range of $r_{\text{ISM}}$ values in Table \ref{tab:lepp}.

From Table \ref{tab:lepp}, we can see that as $T_{\rm iso}$ increases (i.e., going from the low to high GCE setups) the more $^{107}$Pd is underproduced relative to the inferred ESS value. Our results for $f_{\text{Pd}}$ in the low and best GCE setups are consistent with both \cite{trav04} and \cite{bis2014}, wherein the former find that the solar abundances of Sr, Y, and Zr show deficits of $8\%$, $18\%$, and $18\%$, respectively and the latter conclude an additional production channel contributes $\sim25\%$ towards the solar abundances of the $90<A<130$ $s$-isotopes. We note however, that the $^{182}$Hf yield could be up to a factor of two lower than the presently adopted value due to the nuclear uncertainties associated with the $^{182}$Hf/$^{180}$Hf branching point, which means that the $^{107}$Pd/$^{182}$Hf ratio would increase by a factor of two. If this were indeed the case then the predicted $^{107}$Pd/$^{182}$Hf ratio would be higher than $r_{\text{ISM}}$ for all GCE setups. In particular, if we consider the high Monash GCE setup - the only setup for which a self-consistent solution exists for all three radioactive-to-stable ratios assuming that the $^{182}$Hf yield is a factor of two lower (see Section \ref{sec:iso}) - then the GCE predicted  $^{107}$Pd/$^{182}$Hf ratio is $27-50\%$ higher than $r_{\text{ISM}}$. In fact, if the $^{182}$Hf yield is $50\%$ lower than the adopted value, then the predicted $^{107}$Pd/$^{182}$Hf ratio at $t_{\odot}$ is 7.79, which corresponds to a $T_{\text{iso}}$ of $21$ Myr; this is self-consistent with the range of $T_{\text{iso}}$ found in Section \ref{sec:iso} and therefore demonstrates that the possible deficit of the $90<A<130$ $s$-isotopes depends also on the nuclear physics uncertainties associated with the $^{182}$Hf/$^{180}$Hf branching point.

\subsection{Derivation of the time from last event} \label{sec:LE}

The results reported so far apply to the specific case of Regime I of Paper II, when $\tau/\gamma\gtrsim2$, where $\gamma$ is the time interval between the birth of the progenitor of each AGB star that contributed to the composition of the ESS (which is analogous to the average time interval $\delta$ between additions from different AGB stars into the same parcel of ISM gas). Given that $\tau=9.4$ Myr and $12.8$ Myr for $^{107}$Pd and $^{182}$Hf, respectively, Regime I requires that $\gamma\lesssim5$ Myr. However, we also need to consider the possibility that $\gamma$ is much larger than this value since we do not know a priori the value of $\gamma$ for $s$-process AGB sources.

\begin{figure}
    \centering
        \includegraphics[width=1.\linewidth]{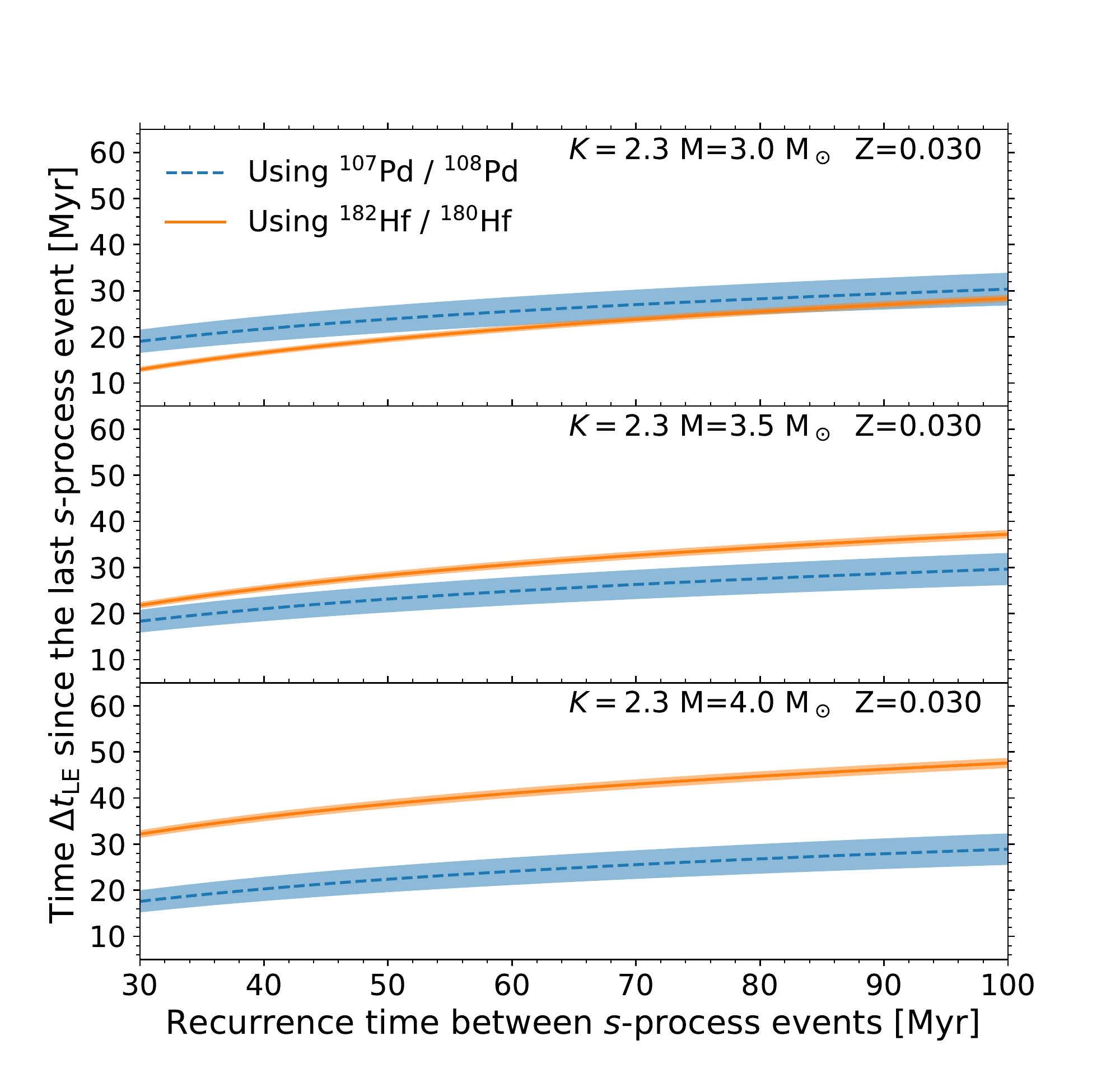}
    \caption{Time elapsed from the last AGB $s$-process event to the formation of the first solids in the ESS as a function of the free parameter $\delta$, the recurrent time interval between $s$-process contributing events. Three stellar AGB models 3, 3.5, and 4 \msun, at $Z=0.03$ are selected as examples, using the best GCE ($K=2.3$). The error bars include the uncertainties on the ESS values and on the mean lives of the two isotopes. For the 3 \msun\ model it is possible to derive several self-consistent solutions, where the blue and the orange bands overlap, depending on the value of the recurrence time ($\delta$) on the x-axis.}
    \label{fig:LE}
\end{figure}

Referring back to the Regimes discussed in Paper II, in particular, if $\tau/\gamma \lesssim 0.3$ (henceforth Regime III), i.e., in the case of $^{107}$Pd and $^{182}$Hf if $\gamma \gtrsim 30$ Myr, the ESS abundances of these SLRs are dominated by the last nucleosynthetic enrichment event that added $s$-process elements to the ESS. In this case we cannot derive a $T_{\text{iso}}$ because the length of time between successive enrichment events is long enough that the SLR can completely decay from the ISM. We can calculate instead the time that elapsed between the last event that added these SLRs to the ESS matter and the formation of the first solids in the ESS, $T_{\rm LE}$. As an example, the $^{107}$Pd/$^{108}$Pd abundance ratio following the last event \citep{wasserburg06,lugaro18} is given by \citep{cote2021},

\begin{equation}
   \frac{^\mathrm{107}\mathrm{Pd}}{^\mathrm{108}\mathrm{Pd}}=K\left(\frac{Y_\mathrm{^{107}Pd}}{Y_\mathrm{^{108}Pd}}\right)\left(\frac{\langle\delta\rangle}{T_\mathrm{gal}}\right),
 \label{eq:LE}
\end{equation}
where $K$ is the GCE parameter described in Paper I, $\delta$ is a free parameter with average time $\gamma \gtrsim 30$ Myr, $T_{\rm gal}$ the age of the Galaxy up to the formation of the Sun (8.4 Gyr), and $Y_\mathrm{^{107}Pd}/Y_\mathrm{^{108}Pd}$ is the production factor in the AGB last event, where we used the stellar yields from Monash models of metallicity $0.007$, $0.01$, $0.014$, and $0.03$ reported in Table~\ref{tab:LE}. For the isotopic ratios of interest here, we needed to add to this estimate the $r$-process component of the stable reference isotopes, \iso{108}Pd and \iso{180}Hf. Using the component from the Monash models as reported in Table~\ref{tab:residual} results in a decrease of the radioactive-to-stable ratio, and therefore of the corresponding $T_{\rm LE}$ by $5 - 9$ Myr and $1.5 - 2$ Myr, when using \iso{107}Pd and \iso{182}Hf, respectively. Table~\ref{tab:LE} shows the models for which it is possible to derive self-consistent $T_{\rm LE}$ using $^{107}$Pd and $^{182}$Hf; Figure~\ref{fig:LE} shows three example cases from Table \ref{tab:LE} for the metallicity $Z=0.03$ of the trend of $T_{\rm LE}$ as a function of the free parameter $\delta$. The top panel shows an example case in which a self-consistent $T_{\rm LE}$ can be found within ESS and mean life uncertainties in the region of $\delta$ where the blue and orange bands overlap. The middle and bottom panel instead show cases where there is no overlap because the $T_{\rm LE}$ derived from $^{182}$Hf is always higher than from $^{107}$Pd. Overall, it is possible for models in the mass range between 2 and 3 \msun\ to obtain self-consistent $T_{\rm LE}$. The upper limit of the ESS $^{135}$Cs/$^{133}$Cs ratio reported in Table~\ref{tab:iso} is also generally consistent with the values of $T_{\rm LE}$. For example, for the 3 \msun, $Z=0.014$ (high $K$) model $T_{\rm LE}$ is $>$ 21 Myr, which is consistent with the interval $29-41$ Myr derived from the other two SLRs. 

When we consider the $^{107}$Pd/$^{182}$Hf ratio, we can find a $T_{\rm LE}$ consistent with the $^{107}$Pd/$^{108}$Pd and $^{107}$Pd/$^{182}$Hf ratios for the 2 \msun, $Z=0.01$ model. To match exactly the $^{107}$Pd/$^{182}$Hf production factor given in Column 5 for this model, we find that the time from the last event to formation of the solids in the ESS is 25.5 Myr. This result is particularly interesting since a 2 \msun\ star is the most common type of AGB star with TDU at this metallicity and they are also the most likely candidates for the parent stars of presolar SiC grains \citep{cristallo20}. Therefore, this model well represents the last AGB source to have added $s$-process elements to the ESS. Instead, at $Z=0.007$, for the 2.1 \msun\ model, the predicted $^{107}$Pd/$^{182}$Hf ratio in Column 5 of Table \ref{tab:LE} is even lower than the ESS ratio. This is because at this metallicity the production of the elements beyond the second $s$-process peak (like Hf) is favoured relative to those between the first and second $s$-process peaks (like Pd). For the higher metallicities, $Z=0.014$ and $Z=0.03$,  we found the reverse problem: the predicted $^{107}$Pd/$^{182}$Hf ratio needs more time to decay to its ESS value than allowed by the $T_{\rm LE}$ calculated on the basis of the radioactive-to-stable ratios. 

In summary, when comparing the time intervals derived using Regime I or Regime III, we find an overall consistency, as expected from our mathematical framework. In fact, $T_{\rm LE}$ is always longer than $T_{\rm iso}$ because the equation used to calculate the radioactive-to-stable ratio after the last event in Regime III (Equation \ref{eq:LE}) differs from the steady-state equation used to calculate the ratio in Regime I (see Equation 1 in Paper I) in that instead of $\tau$ the ratio is proportional to $\delta$. In Regime III by definition $\delta$ is larger than $\tau$, therefore, the ratio is higher and the time is longer. Furthermore, the shorter the $\delta$ the closer is $T_{\rm LE}$ to $T_{\text{iso}}$. The main difference between the two Regimes is that in Regime I we need to invoke an extra source of Pd, while in Regime III we can identify an AGB star of 2 \msun\ and $Z=0.01$ to have potentially been the last to contribute to the Solar System $s$-process elements. 

\begin{table*}
\begin{center}
\caption{Production factors of AGB Monash models of metallicities representative of the metallicity of stars that were present in the solar neighbourhood 4.6 Gyr ago \citep{casagrande11, nissen20}. For four of the eleven models of mass between 2 and 3 \msun\ it is possible to derive a range of self-consistent values of $T_{\rm LE}$, reported in Column 6, with the corresponding Galaxy model (i.e. value of $K$) indicated in brackets. Column 6 reports the \iso{107}Pd/\iso{182}Hf ratio predicted by decaying back the ESS value of 4.25 by the corresponding $T_{\rm LE}$ range, to be compared to the same ratio as derived directly from the models.
\label{tab:LE}}
\begin{tabular}{ccccccc}
\hline

Z & M(\msun) & \iso{107}Pd/\iso{108}Pd & \iso{182}Hf/\iso{180}Hf & \iso{107}Pd/\iso{182}Hf & $T_{\rm LE}$ (Myr) & (\iso{107}Pd/\iso{182}Hf)$_{\rm decay}$ \\ 
 \hline
0.007 & 2.1 & 0.14 & 0.07 & 3.01 & 15 - 22 (low) & 6.49 - 7.91 \\
  &  &  &  & & 19 - 30 (best) &  7.27 - 9.92 \\
    &  &  &  & & 31 - 46 (high) & 10.20 - 15.60  \\
 & 2.5 & 0.14 & 0.11 & 2.45 & \multicolumn{2}{c}{no solution} \\
  & 3 & 0.13 & 0.23 & 1.21 & \multicolumn{2}{c}{no solution} \\ 
                          \hline
0.01 & 2 & 0.14 & 0.05 & 8.73 & 12 - 26 (low) & 5.96 - 8.85 \\
  &  &  &  & & 17 - 31 (best) &  6.87 - 10.20 \\
    &  &  &  & & 26 - 42 (high) & 8.86 - 13.90  \\
& 3 & 0.14 & 0.14 & 5.61 & \multicolumn{2}{c}{no solution} \\
                    \hline
0.014 & 2 & 0.14 & 0.03 & 26.9 & 36 (high) & 11.7 \\ 
& 3 & 0.14 & 0.12 & 7.62 & \multicolumn{2}{c}{no solution} \\
 & 4 & 0.13 & 0.34 & 2.09 & \multicolumn{2}{c}{no solution} \\
 \hline
0.03 & 3 & 0.14 & 0.04 & 27.4 & 17 - 25 (low) & 6.87 - 8.61  \\
  &  &  &  & & 22 - 29 (best) &  7.91 - 9.64  \\
    &  &  &  & & 29 - 41 (high) &  9.64 - 13.5   \\
 & 3.5 & 0.13 & 0.08 & 18.4 & \multicolumn{2}{c}{no solution}  \\
 & 4 & 0.12 & 0.18 & 8.45 & \multicolumn{2}{c}{no solution} \\
                          \hline
\end{tabular}
\end{center}
\end{table*}

\section{Discussion} \label{sec:dis}

For the low, best, and high GCE setups, $T_{\text{iso}}$ is consistent for the Monash and FRUITY yields in the ranges $9-12$ Myr, $10-16$ Myr, and $18-26$ Myr, respectively. Comparing these values of $T_{\text{iso}}$ to those in Paper I (see their Table~2) calculated using $^{107}$Pd and $^{182}$Hf ($^{135}$Cs was not considered), we find that the times here are shorter by more than a factor of two. The main reason for this is that we did not include here an $r$-process source for the SLRs, since the last $r$-process event occurred more than 100 Myr before the formation of the Sun \citep{cote2021}. We only consider the $r$-process component for the stable reference isotopes using the $r$-process residuals. Similarly, in Paper II an $r$-process source for the SLRs was not included, however, constant production ratios (as in Paper I) were used instead of the more realistic stellar nucleosynthesis yields we use in this work. Our isolation times are also shorter than those reported in Paper II (see their Table~4) by roughly 5 Myr in the case of $^{182}$Hf, but agree for $^{107}$Pd. The discrepancy between these results can be attributed to the fact that the $^{182}$Hf/$^{180}$Hf ratio is extremely sensitive to the stellar mass and metallicity, unlike $^{107}$Pd/$^{108}$Pd which mostly depends on the neutron-capture cross section ratio. 

In Section \ref{sec:lepp}, we found that an extra contribution to $^{107}$Pd is needed to recover a $T_{\text{iso}}$ from the $^{107}$Pd/$^{182}$Hf ratio that is consistent with the values obtained using the $^{107}$Pd and $^{182}$Hf radioactive-to-stable abundance ratios. Alternatively, the $^{107}$Pd/$^{182}$Hf ratio in the ISM at $t_{\odot}$ could be increased by reducing the production of $^{182}$Hf via a less efficient activation of the $^{181}$Hf branching point. However, this second solution would remove the current agreement on the $T_{\text{iso}}$ derived using the $^{107}$Pd/$^{108}$Pd and $^{182}$Hf/$^{180}$Hf ratios with the Monash models. Therefore, we need to invoke an extra source for the production of the first $s$-peak elements like $^{107}$Pd in the Solar System. An additional process, often referred to as the lighter element primary process (LEPP) has been postulated to play a role in the production of isotopes at and around the first $s$-peak \citep{trav04, bis2014, bis17} and the amount of $^{107}$Pd that is ``missing'' in our low and best Monash GCE setups is consistent with the LEPP contribution invoked by both \cite{trav04} and \cite{bis2014}. 

The LEPP and the different nucleosynthesis processes possibly contributing to the process have been a matter of discussion \citep[see e.g.][and references therein]{montes07, qian:07, farouqi:10, arc011}. Also, the need of a LEPP to reproduce the solar abundances has been questioned, particularly when considering the FRUITY models \citep[e.g.][]{cristallo2015b}. Indeed our GCE models show that the FRUITY models produce more first peak $s$-process elements than the Monash models (Figure~\ref{fig:prod_factor}), however, in this case $^{208}$Pb is underproduced. Using FRUITY AGB yields, \cite{pra18, pra20} were able to resolve this issue by including in their GCE model rotating massive star yields from \cite{limongi2018} which provide an additional contribution to the first peak $s$-isotopes. Since the GCE models of \cite{pra18, pra20} also reproduce well the third $s$-process peak (including $^{208}$Pb), this suggests that the existence of a possible deficit of the first $s$-process peak may be a consequence of the choice of yields as well as the GCE model. However, it must be considered that like AGB stellar yields, the yields for rotating massive stars also have uncertainties and the significant variations obtained between different sets of models are not surprising. Among other things, this may be due to the different approaches used to implement the rotation mechanism in one-dimensional models and to the efficiency of rotation in affecting the stellar structure at different metallicities. In \cite{bri21} the authors show that on the lower mass-end of massive stars ($10-35$ \msun), the rotating models of \cite{limongi2018} show features that are not present in other rotating massive star yields. On the other hand, \cite{riz19} found that their GCE model can best reproduce the observed $s$-process abundances in the Milky Way when using the rotating massive star yields of \cite{fri16} or, in the case of the \cite{limongi2018} yields, if they assume that only the stars at the lowest metallicity slowly rotate. The uncertainties coming from GCE, including the different assumptions adopted by \cite{riz19} and \cite{pra18, pra20} concerning the metallicity dependence of the rotational velocity distribution for rotating massive stars, must be taken into account mean that it not yet possible to derive effective constraints between different stellar sets. Additionally, as discussed by \cite{pig08} and \cite{fri16}, the $s$-process production in rotating massive stars is highly affected by nuclear uncertainties (e.g., by $\alpha$ capture rates on $^{22}$Ne and $^{17}$O). It is evident that the effective contribution of fast rotating massive stars to GCE is still a matter of debate, and therefore we believe that the existence of a LEPP is still an open question. 

A potential solution to the missing $^{107}$Pd may be found by considering an enhanced contribution to the solar abundances of the $s$-process elements from AGB stars of a higher metallicity than those that were assumed to contribute to the ESS in this work. For example, the effect of stellar migration could have moved higher metallicity AGB stars from the inner region of the Galaxy to the location of the formation of the Sun \citep{wielen1996, minchev2013, minchev2014, kub15, cristallo20}. These stars could potentially have increased the abundances of the first $s$-peak elements and of s-process isotopes in the mass region between Sr and the second $s$-process peak in the ESS, without contributing any additional iron.

We cannot model this processes in our code, but to test if this solution would work qualitatively we calibrated a best fit GCE setup using the Monash yields that reached $Z=0.018$ (instead of 0.014) at $t_{\odot}$. The isolation times derived from this test model using the radioactive-to-stable ratios were of the order of 12 Myr, and the $^{107}$Pd/$^{182}$Hf ratio at $t_{\odot}$ is 7.8 which, when decayed to the ESS value, results in an isolation time of 30 Myr. Therefore, in principle it would be possible to find a self-consistent solution with a model run calibrating the solar $Z$ somewhere in-between 0.014 and 0.018. This alternative solution for the LEPP in the Solar System needs to be investigated with more sophisticated GCE models, considering the balance between the first and third $s$-process peak, and the fact that metal-rich AGB stars may contribute to the chemical enrichment history of the Sun. Furthermore, the abundances of all elements between the first and the second $s$-process peaks will need to be reproduced consistently.

Overall, our analysis can provide new, independent, accurate, and precise constraints in the form of the ESS $^{107}$Pd/$^{182}$Hf ratio to the open question of the production of the first $s$-process peak elements in the Milky Way disk. However, before we can make robust conclusions it will be necessary to analyse the $^{107}$Pd/$^{182}$Hf ratio also using FRUITY yields, but with an updated $^{181}$Hf decay rate. Since our results only apply to the solar abundances, however, they cannot be used to infer whether a missing contribution of the first peak $s$-isotopes was already active in the early Galaxy.

\section{Conclusion}

In this work, we investigated the origin of $^{107}$Pd, $^{135}$Cs, and $^{182}$Hf in the ESS using the \texttt{OMEGA+} GCE code. We calculate timescales relevant for the birth of the Sun by comparing our predicted radioactive-to-stable abundance ratios in the ISM at $t_{\odot}$ to isotopic abundance ratios in primitive meteorites. We simulate three Milky Way setups for each of two sets of mass and metallicity dependent theoretical AGB yields (Monash and FRUITY), so that our timescales account for uncertainties due to GCE and stellar nucleosynthesis modelling. At $t_{\odot}$, the uncertainty factors for the abundance of an SLR in the ISM depends on the ratio of its mean-life, $\tau$, to the average length of time between the birth of the progenitor of the AGB stars that contributed to the ESS, $\gamma$. Since the latter is poorly constrained, we calculate timescales for two different cases of $\tau/\gamma$. The main results are:

\begin{itemize}
    \item {If $\tau/\gamma \gtrsim2$ (Regime I), we calculate an isolation time, $T_{\text{iso}}$, between 9 and 26 Myr. This range is self-consistent for all radioactive-to-stable abundance ratios investigated in this work and takes into account a $1\sigma$ uncertainty due to the effects of ISM heterogeneities on the radioactive-to-stable ratio at $t_{\odot}$.}
    
    \item {Assuming that we are in Regime I, the predicted $^{107}$Pd/$^{182}$Hf ratio indicates that $9-73\%$ of $^{107,108}$Pd in the ESS is missing from our GCE setups. If the nuclear physics inputs we used here will be confirmed by future experiments and theory, we postulate two potential solutions to this problem: $(1)$ an additional stellar production mechanism for the first $s$-peak isotopes, such as a \textit{lighter element primary process}; or $(2)$ an enhanced contribution of the first $s$-peak isotopes to the solar abundances from higher metallicity stars that migrated from the inner disk of the Galaxy. We find also a solution in the high GCE setup by considering the nuclear physics uncertainties associated with the $^{182}$Hf/$^{180}$Hf branching point.}
    
    \item {If $\tau/\gamma \lesssim 0.3$ (Regime III), we find that for the Monash models in the range $2-3$ \msun\ at $Z=0.007$, $0.014$, and $0.03$ the time from the last event, $T_{\text{LE}}$, is self-consistent for all radioactive-to-stable ratios explored in this work. Furthermore, we identify a single Monash model (2 \msun, $Z=0.01$) for which $T_{\rm LE}=25.5$ Myr is a self-consistent solution which takes into account also the $^{107}$Pd/$^{182}$Hf ratio. Importantly, this solution exists without the need to invoke an extra Pd source in the Galaxy.}
\end{itemize}

The methodology presented in this work can be used to follow the evolution of the radioactive-to-stable abundance ratio for any SLR in the ISM, for which stellar nucleosynthesis yields for a range of masses and metallicity are available, in order to better understand the birth environment of the Sun. In future work we would like to include the yields from the rotating massive star models of \cite{limongi2018} in our GCE framework, to see whether we reach the same conclusion as \cite{pra18, pra20}: that an additional contribution of the $A<90$ $s$-isotopes in the ESS is not needed, owing to the increased weak $s$-process in rotating massive star models. Also, a better treatment of the $r$-process sources in the Galaxy in our calculations should be implemented and models of transport of radioactive nuclei in the ISM are needed to better asses the value of $\gamma$ for the $s$-process.

Furthermore, as we have shown here, our approach can help us clarify the production mechanism and stellar sources in the Galaxy, and it could also be applied to constrain nuclear physic properties. An example of this is the case of $^{205}$Pb (with a $\tau$ of 25 Myr), which is produced by the $s$-process in AGB stars and known to be present in the ESS (although the evidence is weak and awaits confirmation). However, the isotope's electron-capture rate is strongly temperature and density dependent and its variations are not well determined in stellar environments \citep{mow98}. Future work could consider this isotope and constrain its nuclear properties by comparing $T_{\text{iso}}$ derived using the $^{205}$Pb/$^{204}$Pb ratio to those derived here.

\section{Acknowledgements}

We are thankful to the referee for helping to improve the manuscript through their constructive comments and suggestions. We thank also Blanka Vil\'{a}gos for helping to test several important reaction rates using the Monash models and Claudia Travaglio for discussion. This research is supported by the ERC Consolidator Grant (Hungary) programme (Project RADIOSTAR, G.A. n. 724560).  We thank the ChETEC COST Action (CA16117), supported by the European Cooperation in Science and Technology, the ChETEC-INFRA project funded from the European Union’s Horizon 2020 research and innovation programme under grant agreement No 101008324, and the IReNA network supported by NSF AccelNet. TT and MP acknowledge significant support to NuGrid from STFC (through the University of Hull's Consolidated Grant ST/R000840/1) and ongoing access to {\tt viper}, the University of Hull High Performance Computing Facility. BC and MP are grateful for support from the National Science Foundation (NSF, USA) under grant No. PHY-1430152 (JINA Center for the Evolution of the Elements), and from the "Lendulet-2014" Program of the Hungarian Academy of Sciences (Hungary). AK was supported by the Australian Research Council Centre of Excellence for All Sky Astrophysics in 3 Dimensions (ASTRO 3D), through project number CE170100013. 

\bibliography{slrs}{}

\begin{thebibliography}{}
\expandafter\ifx\csname natexlab\endcsname\relax\def\natexlab#1{#1}\fi
\providecommand{\url}[1]{\href{#1}{#1}}
\providecommand{\dodoi}[1]{doi:~\href{http://doi.org/#1}{\nolinkurl{#1}}}
\providecommand{\doeprint}[1]{\href{http://ascl.net/#1}{\nolinkurl{http://ascl.net/#1}}}
\providecommand{\doarXiv}[1]{\href{https://arxiv.org/abs/#1}{\nolinkurl{https://arxiv.org/abs/#1}}}

\bibitem[{{Arcones} \& {Montes}(2011)}]{arc011}
{Arcones}, A., \& {Montes}, F. 2011, \apj, 731, 5,
  \dodoi{10.1088/0004-637X/731/1/5}

\bibitem[{{Arlandini} {et~al.}(1999){Arlandini}, {K{\"a}ppeler}, {Wisshak},
  {Gallino}, {Lugaro}, {Busso}, \& {Straniero}}]{arlandini99}
{Arlandini}, C., {K{\"a}ppeler}, F., {Wisshak}, K., {et~al.} 1999, \apj, 525,
  886, \dodoi{10.1086/307938}

\bibitem[{{Asplund} {et~al.}(2009){Asplund}, {Grevesse}, {Sauval}, \&
  {Scott}}]{asp09}
{Asplund}, M., {Grevesse}, N., {Sauval}, A.~J., \& {Scott}, P. 2009, \araa, 47,
  481, \dodoi{10.1146/annurev.astro.46.060407.145222}

\bibitem[{{Battino} {et~al.}(2021){Battino}, {Lederer-Woods}, {Cseh},
  {Denissenkov}, \& {Herwig}}]{bat21}
{Battino}, U., {Lederer-Woods}, C., {Cseh}, B., {Denissenkov}, P., \& {Herwig},
  F. 2021, Universe, 7, 25, \dodoi{10.3390/universe7020025}

\bibitem[{{Battino} {et~al.}(2016){Battino}, {Pignatari}, {Ritter}, {Herwig},
  {Denisenkov}, {Den Hartogh}, {Trappitsch}, {Hirschi}, {Freytag},
  {Thielemann}, \& {Paxton}}]{bat16}
{Battino}, U., {Pignatari}, M., {Ritter}, C., {et~al.} 2016, \apj, 827, 30,
  \dodoi{10.3847/0004-637X/827/1/30}

\bibitem[{{Battino} {et~al.}(2019){Battino}, {Tattersall}, {Lederer-Woods},
  {Herwig}, {Denissenkov}, {Hirschi}, {Trappitsch}, {den Hartogh}, {Pignatari},
  \& {NuGrid Collaboration}}]{bat19}
{Battino}, U., {Tattersall}, A., {Lederer-Woods}, C., {et~al.} 2019, \mnras,
  489, 1082, \dodoi{10.1093/mnras/stz2158}

\bibitem[{{Bisterzo} {et~al.}(2014){Bisterzo}, {Travaglio}, {Gallino},
  {Wiescher}, \& {K{\"a}ppeler}}]{bis2014}
{Bisterzo}, S., {Travaglio}, C., {Gallino}, R., {Wiescher}, M., \&
  {K{\"a}ppeler}, F. 2014, \apj, 787, 10, \dodoi{10.1088/0004-637X/787/1/10}

\bibitem[{{Bisterzo} {et~al.}(2017){Bisterzo}, {Travaglio}, {Wiescher},
  {K{\"a}ppeler}, \& {Gallino}}]{bis17}
{Bisterzo}, S., {Travaglio}, C., {Wiescher}, M., {K{\"a}ppeler}, F., \&
  {Gallino}, R. 2017, \apj, 835, 97, \dodoi{10.3847/1538-4357/835/1/97}

\bibitem[{{Bisterzo} {et~al.}(2015){Bisterzo}, {Gallino}, {K{\"a}ppeler},
  {Wiescher}, {Imbriani}, {Straniero}, {Cristallo}, {G{\"o}rres}, \&
  {deBoer}}]{bis15}
{Bisterzo}, S., {Gallino}, R., {K{\"a}ppeler}, F., {et~al.} 2015, \mnras, 449,
  506, \dodoi{10.1093/mnras/stv271}

\bibitem[{{Bondarenko} {et~al.}(2002){Bondarenko}, {Berzins}, {Prokofjevs},
  {Simonova}, {von Egidy}, {Honz{\'a}tko}, {Tomand l}, {Alexa}, {Wirth},
  {K{\"o}ster}, {Eisermann}, {Metz}, {Graw}, {Hertenberger}, \&
  {Rubacek}}]{bon02}
{Bondarenko}, V., {Berzins}, J., {Prokofjevs}, P., {et~al.} 2002, \nphysa, 709,
  3, \dodoi{10.1016/S0375-9474(02)00646-2}

\bibitem[{{Brinkman} {et~al.}(2021){Brinkman}, {den Hartogh}, {Doherty},
  {Pignatari}, \& {Lugaro}}]{bri21}
{Brinkman}, H.~E., {den Hartogh}, J.~W., {Doherty}, C.~L., {Pignatari}, M., \&
  {Lugaro}, M. 2021, arXiv e-prints, arXiv:2109.05842.
\newblock \doarXiv{2109.05842}

\bibitem[{{Buntain} {et~al.}(2017){Buntain}, {Doherty}, {Lugaro}, {Lattanzio},
  {Stancliffe}, \& {Karakas}}]{bun17}
{Buntain}, J.~F., {Doherty}, C.~L., {Lugaro}, M., {et~al.} 2017, \mnras, 471,
  824, \dodoi{10.1093/mnras/stx1502}

\bibitem[{{Burbidge} {et~al.}(1957){Burbidge}, {Burbidge}, {Fowler}, \&
  {Hoyle}}]{burbidge57}
{Burbidge}, E.~M., {Burbidge}, G.~R., {Fowler}, W.~A., \& {Hoyle}, F. 1957,
  Reviews of Modern Physics, 29, 547, \dodoi{10.1103/RevModPhys.29.547}

\bibitem[{{Busso} {et~al.}(1999){Busso}, {Gallino}, \& {Wasserburg}}]{busso99}
{Busso}, M., {Gallino}, R., \& {Wasserburg}, G.~J. 1999, \araa, 37, 239,
  \dodoi{10.1146/annurev.astro.37.1.239}

\bibitem[{{Busso} {et~al.}(2020){Busso}, {Vescovi}, {Palmerini}, {Cristallo},
  \& {Antonuccio Delogu}}]{bus20}
{Busso}, M., {Vescovi}, D., {Palmerini}, S., {Cristallo}, S., \& {Antonuccio
  Delogu}, V. 2020, arXiv e-prints, arXiv:2011.07469.
\newblock \doarXiv{2011.07469}

\bibitem[{{Casagrande} {et~al.}(2011){Casagrande}, {Sch{\"o}nrich}, {Asplund},
  {Cassisi}, {Ram{\'\i}rez}, {Mel{\'e}ndez}, {Bensby}, \&
  {Feltzing}}]{casagrande11}
{Casagrande}, L., {Sch{\"o}nrich}, R., {Asplund}, M., {et~al.} 2011, \aap, 530,
  A138, \dodoi{10.1051/0004-6361/201016276}

\bibitem[{{Chiappini} {et~al.}(1997){Chiappini}, {Matteucci}, \&
  {Gratton}}]{chi97}
{Chiappini}, C., {Matteucci}, F., \& {Gratton}, R. 1997, \apj, 477, 765,
  \dodoi{10.1086/303726}

\bibitem[{{C{\^o}t{\'e}} {et~al.}(2019{\natexlab{a}}){C{\^o}t{\'e}}, {Lugaro},
  {Reifarth}, {Pignatari}, {Vil{\'a}gos}, {Yag{\"u}e}, \& {Gibson}}]{cote19}
{C{\^o}t{\'e}}, B., {Lugaro}, M., {Reifarth}, R., {et~al.} 2019{\natexlab{a}},
  \apj, 878, 156, \dodoi{10.3847/1538-4357/ab21d1}

\bibitem[{{C{\^o}t{\'e}} {et~al.}(2017){C{\^o}t{\'e}}, {O'Shea}, {Ritter},
  {Herwig}, \& {Venn}}]{cote17}
{C{\^o}t{\'e}}, B., {O'Shea}, B.~W., {Ritter}, C., {Herwig}, F., \& {Venn},
  K.~A. 2017, \apj, 835, 128, \dodoi{10.3847/1538-4357/835/2/128}

\bibitem[{{C{\^o}t{\'e}} {et~al.}(2018){C{\^o}t{\'e}}, {Silvia}, {O'Shea},
  {Smith}, \& {Wise}}]{cote18}
{C{\^o}t{\'e}}, B., {Silvia}, D.~W., {O'Shea}, B.~W., {Smith}, B., \& {Wise},
  J.~H. 2018, \apj, 859, 67, \dodoi{10.3847/1538-4357/aabe8f}

\bibitem[{{C{\^o}t{\'e}} {et~al.}(2019{\natexlab{b}}){C{\^o}t{\'e}},
  {Yag{\"u}e}, {Vil{\'a}gos}, \& {Lugaro}}]{cote19b}
{C{\^o}t{\'e}}, B., {Yag{\"u}e}, A., {Vil{\'a}gos}, B., \& {Lugaro}, M.
  2019{\natexlab{b}}, arXiv e-prints, arXiv:1911.01457.
\newblock \doarXiv{1911.01457}

\bibitem[{{C{\^o}t{\'e}} {et~al.}(2021){C{\^o}t{\'e}}, {Eichler}, {Yag{\"u}e
  L{\'o}pez}, {Vassh}, {Mumpower}, {Vil{\'a}gos}, {So{\'o}s}, {Arcones},
  {Sprouse}, {Surman}, {Pignatari}, {Pet{\H{o}}}, {Wehmeyer}, {Rauscher}, \&
  {Lugaro}}]{cote2021}
{C{\^o}t{\'e}}, B., {Eichler}, M., {Yag{\"u}e L{\'o}pez}, A., {et~al.} 2021,
  Science, 371, 945, \dodoi{10.1126/science.aba1111}

\bibitem[{{Cristallo} {et~al.}(2015{\natexlab{a}}){Cristallo}, {Abia},
  {Straniero}, \& {Piersanti}}]{cristallo2015b}
{Cristallo}, S., {Abia}, C., {Straniero}, O., \& {Piersanti}, L.
  2015{\natexlab{a}}, \apj, 801, 53, \dodoi{10.1088/0004-637X/801/1/53}

\bibitem[{{Cristallo} {et~al.}(2020){Cristallo}, {Nanni}, {Cescutti},
  {Minchev}, {Liu}, {Vescovi}, {Gobrecht}, \& {Piersanti}}]{cristallo20}
{Cristallo}, S., {Nanni}, A., {Cescutti}, G., {et~al.} 2020, \aap, 644, A8,
  \dodoi{10.1051/0004-6361/202039492}

\bibitem[{{Cristallo} {et~al.}(2009){Cristallo}, {Straniero}, {Gallino},
  {Piersanti}, {Dom{\'\i}nguez}, \& {Lederer}}]{cristallo09}
{Cristallo}, S., {Straniero}, O., {Gallino}, R., {et~al.} 2009, \apj, 696, 797,
  \dodoi{10.1088/0004-637X/696/1/797}

\bibitem[{{Cristallo} {et~al.}(2008){Cristallo}, {Straniero}, \&
  {Lederer}}]{cris08}
{Cristallo}, S., {Straniero}, O., \& {Lederer}, M.~T. 2008, in American
  Institute of Physics Conference Series, Vol. 990, First Stars III, ed. B.~W.
  {O'Shea} \& A.~{Heger}, 320--324, \dodoi{10.1063/1.2905571}

\bibitem[{{Cristallo} {et~al.}(2015{\natexlab{b}}){Cristallo}, {Straniero},
  {Piersanti}, \& {Gobrecht}}]{cris15}
{Cristallo}, S., {Straniero}, O., {Piersanti}, L., \& {Gobrecht}, D.
  2015{\natexlab{b}}, \apjs, 219, 40, \dodoi{10.1088/0067-0049/219/2/40}

\bibitem[{{Cristallo} {et~al.}(2011){Cristallo}, {Piersanti}, {Straniero},
  {Gallino}, {Dom{\'\i}nguez}, {Abia}, {Di Rico}, {Quintini}, \&
  {Bisterzo}}]{cris11}
{Cristallo}, S., {Piersanti}, L., {Straniero}, O., {et~al.} 2011, \apjs, 197,
  17, \dodoi{10.1088/0067-0049/197/2/17}

\bibitem[{{Cseh} {et~al.}(2018){Cseh}, {Lugaro}, {D'Orazi}, {de Castro},
  {Pereira}, {Karakas}, {Moln{\'a}r}, {Plachy}, {Szab{\'o}}, {Pignatari}, \&
  {Cristallo}}]{cseh18}
{Cseh}, B., {Lugaro}, M., {D'Orazi}, V., {et~al.} 2018, \aap, 620, A146,
  \dodoi{10.1051/0004-6361/201834079}

\bibitem[{{Dauphas} \& {Chaussidon}(2011)}]{dauphas}
{Dauphas}, N., \& {Chaussidon}, M. 2011, Annual Review of Earth and Planetary
  Sciences, 39, 351, \dodoi{10.1146/annurev-earth-040610-133428}

\bibitem[{{den Hartogh} {et~al.}(2019){den Hartogh}, {Hirschi}, {Lugaro},
  {Doherty}, {Battino}, {Herwig}, {Pignatari}, \& {Eggenberger}}]{denhartogh}
{den Hartogh}, J.~W., {Hirschi}, R., {Lugaro}, M., {et~al.} 2019, \aap, 629,
  A123, \dodoi{10.1051/0004-6361/201935476}

\bibitem[{{Denissenkov} \& {Tout}(2003)}]{den03}
{Denissenkov}, P.~A., \& {Tout}, C.~A. 2003, \mnras, 340, 722,
  \dodoi{10.1046/j.1365-8711.2003.06284.x}

\bibitem[{{Farouqi} {et~al.}(2010){Farouqi}, {Kratz}, {Pfeiffer}, {Rauscher},
  {Thielemann}, \& {Truran}}]{farouqi:10}
{Farouqi}, K., {Kratz}, K.~L., {Pfeiffer}, B., {et~al.} 2010, \apj, 712, 1359,
  \dodoi{10.1088/0004-637X/712/2/1359}

\bibitem[{{Firestone}(1991)}]{firestone91}
{Firestone}, R.~B. 1991, Nuclear Data Sheets, 62, 101,103,
  \dodoi{10.1016/0090-3752(91)80014-W}

\bibitem[{{Fishlock} {et~al.}(2014){Fishlock}, {Karakas}, {Lugaro}, \&
  {Yong}}]{fish14}
{Fishlock}, C.~K., {Karakas}, A.~I., {Lugaro}, M., \& {Yong}, D. 2014, \apj,
  797, 44, \dodoi{10.1088/0004-637X/797/1/44}

\bibitem[{{Frischknecht} {et~al.}(2016){Frischknecht}, {Hirschi}, {Pignatari},
  {Maeder}, {Meynet}, {Chiappini}, {Thielemann}, {Rauscher}, {Georgy}, \&
  {Ekstr{\"o}m}}]{fri16}
{Frischknecht}, U., {Hirschi}, R., {Pignatari}, M., {et~al.} 2016, \mnras, 456,
  1803, \dodoi{10.1093/mnras/stv2723}

\bibitem[{{Gallino} {et~al.}(1998){Gallino}, {Arlandini}, {Busso}, {Lugaro},
  {Travaglio}, {Straniero}, {Chieffi}, \& {Limongi}}]{gal98}
{Gallino}, R., {Arlandini}, C., {Busso}, M., {et~al.} 1998, \apj, 497, 388,
  \dodoi{10.1086/305437}

\bibitem[{{Goriely} \& {Mowlavi}(2000)}]{gor00}
{Goriely}, S., \& {Mowlavi}, N. 2000, \aap, 362, 599

\bibitem[{{Goriely} \& {Siess}(2004)}]{gor04}
{Goriely}, S., \& {Siess}, L. 2004, \aap, 421, L25,
  \dodoi{10.1051/0004-6361:20040184}

\bibitem[{{Herwig}(2005)}]{her05}
{Herwig}, F. 2005, \araa, 43, 435,
  \dodoi{10.1146/annurev.astro.43.072103.150600}

\bibitem[{{Herwig} {et~al.}(1997){Herwig}, {Bloecker}, {Schoenberner}, \& {El
  Eid}}]{her97}
{Herwig}, F., {Bloecker}, T., {Schoenberner}, D., \& {El Eid}, M. 1997, \aap,
  324, L81.
\newblock \doarXiv{astro-ph/9706122}

\bibitem[{{Huss} {et~al.}(2009){Huss}, {Meyer}, {Srinivasan}, {Goswami}, \&
  {Sahijpal}}]{huss09}
{Huss}, G.~R., {Meyer}, B.~S., {Srinivasan}, G., {Goswami}, J.~N., \&
  {Sahijpal}, S. 2009, \gca, 73, 4922, \dodoi{10.1016/j.gca.2009.01.039}

\bibitem[{{Jones} {et~al.}(2019){Jones}, {C{\^o}t{\'e}}, {R{\"o}pke}, \&
  {Wanajo}}]{jones19}
{Jones}, S., {C{\^o}t{\'e}}, B., {R{\"o}pke}, F.~K., \& {Wanajo}, S. 2019,
  \apj, 882, 170, \dodoi{10.3847/1538-4357/ab384e}

\bibitem[{{K{\"a}ppeler} {et~al.}(2011){K{\"a}ppeler}, {Gallino}, {Bisterzo},
  \& {Aoki}}]{kap11}
{K{\"a}ppeler}, F., {Gallino}, R., {Bisterzo}, S., \& {Aoki}, W. 2011, Reviews
  of Modern Physics, 83, 157, \dodoi{10.1103/RevModPhys.83.157}

\bibitem[{{Karakas} {et~al.}(2012){Karakas}, {Garc{\'\i}a-Hern{\'a}ndez}, \&
  {Lugaro}}]{kar12}
{Karakas}, A.~I., {Garc{\'\i}a-Hern{\'a}ndez}, D.~A., \& {Lugaro}, M. 2012,
  \apj, 751, 8, \dodoi{10.1088/0004-637X/751/1/8}

\bibitem[{{Karakas} \& {Lattanzio}(2014)}]{karakas2014}
{Karakas}, A.~I., \& {Lattanzio}, J.~C. 2014, \pasa, 31, e030,
  \dodoi{10.1017/pasa.2014.21}

\bibitem[{{Karakas} \& {Lugaro}(2016)}]{kar16}
{Karakas}, A.~I., \& {Lugaro}, M. 2016, \apj, 825, 26,
  \dodoi{10.3847/0004-637X/825/1/26}

\bibitem[{{Karakas} {et~al.}(2018){Karakas}, {Lugaro}, {Carlos}, {Cseh},
  {Kamath}, \& {Garc{\'\i}a-Hern{\'a}ndez}}]{kar18}
{Karakas}, A.~I., {Lugaro}, M., {Carlos}, M., {et~al.} 2018, \mnras, 477, 421,
  \dodoi{10.1093/mnras/sty625}

\bibitem[{{Kastner} \& {Myers}(1994)}]{kas94}
{Kastner}, J.~H., \& {Myers}, P.~C. 1994, \apj, 421, 605,
  \dodoi{10.1086/173676}

\bibitem[{{Kubryk} {et~al.}(2015){Kubryk}, {Prantzos}, \&
  {Athanassoula}}]{kub15}
{Kubryk}, M., {Prantzos}, N., \& {Athanassoula}, E. 2015, \aap, 580, A126,
  \dodoi{10.1051/0004-6361/201424171}

\bibitem[{{Lichtenberg} {et~al.}(2016){Lichtenberg}, {Golabek}, {Gerya}, \&
  {Meyer}}]{lichtenberg}
{Lichtenberg}, T., {Golabek}, G.~J., {Gerya}, T.~V., \& {Meyer}, M.~R. 2016,
  \icarus, 274, 350, \dodoi{10.1016/j.icarus.2016.03.004}

\bibitem[{{Limongi} \& {Chieffi}(2018)}]{limongi2018}
{Limongi}, M., \& {Chieffi}, A. 2018, \apjs, 237, 13,
  \dodoi{10.3847/1538-4365/aacb24}

\bibitem[{{Lodders} {et~al.}(2009){Lodders}, {Palme}, \& {Gail}}]{lod09}
{Lodders}, K., {Palme}, H., \& {Gail}, H.~P. 2009, Landolt B\&ouml;rnstein, 4B,
  712, \dodoi{10.1007/978-3-540-88055-4_34}

\bibitem[{{Longland} {et~al.}(2010){Longland}, {Iliadis}, {Champagne},
  {Newton}, {Ugalde}, {Coc}, \& {Fitzgerald}}]{lon10}
{Longland}, R., {Iliadis}, C., {Champagne}, A.~E., {et~al.} 2010, \nphysa, 841,
  1, \dodoi{10.1016/j.nuclphysa.2010.04.008}

\bibitem[{{Lugaro} {et~al.}(2012){Lugaro}, {Doherty}, {Karakas}, {Maddison},
  {Liffman}, {Garc{\'\i}a-Hern{\'a}ndez}, {Siess}, \& {Lattanzio}}]{lugaro12}
{Lugaro}, M., {Doherty}, C.~L., {Karakas}, A.~I., {et~al.} 2012, Meteoritics
  and Planetary Science, 47, 1998, \dodoi{10.1111/j.1945-5100.2012.01411.x}

\bibitem[{{Lugaro} {et~al.}(2018){Lugaro}, {Ott}, \& {Kereszturi}}]{lugaro18}
{Lugaro}, M., {Ott}, U., \& {Kereszturi}, {\'A}. 2018, Progress in Particle and
  Nuclear Physics, 102, 1, \dodoi{10.1016/j.ppnp.2018.05.002}

\bibitem[{{Lugaro} {et~al.}(2014){Lugaro}, {Heger}, {Osrin}, {Goriely},
  {Zuber}, {Karakas}, {Gibson}, {Doherty}, {Lattanzio}, \& {Ott}}]{lugaro14}
{Lugaro}, M., {Heger}, A., {Osrin}, D., {et~al.} 2014, Science, 345, 650,
  \dodoi{10.1126/science.1253338}

\bibitem[{{Meyer} \& {Clayton}(2000)}]{meyer00}
{Meyer}, B.~S., \& {Clayton}, D.~D. 2000, \ssr, 92, 133,
  \dodoi{10.1023/A:1005282825778}

\bibitem[{{Minchev} {et~al.}(2013){Minchev}, {Chiappini}, \&
  {Martig}}]{minchev2013}
{Minchev}, I., {Chiappini}, C., \& {Martig}, M. 2013, \aap, 558, A9,
  \dodoi{10.1051/0004-6361/201220189}

\bibitem[{{Minchev} {et~al.}(2014){Minchev}, {Chiappini}, \&
  {Martig}}]{minchev2014}
---. 2014, \aap, 572, A92, \dodoi{10.1051/0004-6361/201423487}

\bibitem[{{Montes} {et~al.}(2007){Montes}, {Beers}, {Cowan}, {Elliot},
  {Farouqi}, {Gallino}, {Heil}, {Kratz}, {Pfeiffer}, {Pignatari}, \&
  {Schatz}}]{montes07}
{Montes}, F., {Beers}, T.~C., {Cowan}, J., {et~al.} 2007, \apj, 671, 1685,
  \dodoi{10.1086/523084}

\bibitem[{{Mowlavi} {et~al.}(1998){Mowlavi}, {Goriely}, \& {Arnould}}]{mow98}
{Mowlavi}, N., {Goriely}, S., \& {Arnould}, M. 1998, \aap, 330, 206.
\newblock \doarXiv{astro-ph/9711025}

\bibitem[{{Nissen} {et~al.}(2020){Nissen}, {Christensen-Dalsgaard},
  {Mosumgaard}, {Silva Aguirre}, {Spitoni}, \& {Verma}}]{nissen20}
{Nissen}, P.~E., {Christensen-Dalsgaard}, J., {Mosumgaard}, J.~R., {et~al.}
  2020, \aap, 640, A81, \dodoi{10.1051/0004-6361/202038300}

\bibitem[{{Ott} \& {Kratz}(2008)}]{ott08}
{Ott}, U., \& {Kratz}, K.-L. 2008, \nar, 52, 396,
  \dodoi{10.1016/j.newar.2008.05.001}

\bibitem[{{Pignatari} {et~al.}(2010){Pignatari}, {Gallino}, {Heil}, {Wiescher},
  {K{\"a}ppeler}, {Herwig}, \& {Bisterzo}}]{pignatari:2010}
{Pignatari}, M., {Gallino}, R., {Heil}, M., {et~al.} 2010, \apj, 710, 1557,
  \dodoi{10.1088/0004-637X/710/2/1557}

\bibitem[{{Pignatari} {et~al.}(2008){Pignatari}, {Gallino}, {Meynet},
  {Hirschi}, {Herwig}, \& {Wiescher}}]{pig08}
{Pignatari}, M., {Gallino}, R., {Meynet}, G., {et~al.} 2008, \apjl, 687, L95,
  \dodoi{10.1086/593350}

\bibitem[{{Pignatari} {et~al.}(2016){Pignatari}, {Herwig}, {Hirschi},
  {Bennett}, {Rockefeller}, {Fryer}, {Timmes}, {Ritter}, {Heger}, {Jones},
  {Battino}, {Dotter}, {Trappitsch}, {Diehl}, {Frischknecht}, {Hungerford},
  {Magkotsios}, {Travaglio}, \& {Young}}]{nugrid1}
{Pignatari}, M., {Herwig}, F., {Hirschi}, R., {et~al.} 2016, \apjs, 225, 24,
  \dodoi{10.3847/0067-0049/225/2/24}

\bibitem[{{Prantzos} {et~al.}(2020){Prantzos}, {Abia}, {Cristallo}, {Limongi},
  \& {Chieffi}}]{pra20}
{Prantzos}, N., {Abia}, C., {Cristallo}, S., {Limongi}, M., \& {Chieffi}, A.
  2020, \mnras, 491, 1832, \dodoi{10.1093/mnras/stz3154}

\bibitem[{{Prantzos} {et~al.}(2018){Prantzos}, {Abia}, {Limongi}, {Chieffi}, \&
  {Cristallo}}]{pra18}
{Prantzos}, N., {Abia}, C., {Limongi}, M., {Chieffi}, A., \& {Cristallo}, S.
  2018, \mnras, 476, 3432, \dodoi{10.1093/mnras/sty316}

\bibitem[{{Qian} \& {Wasserburg}(2007)}]{qian:07}
{Qian}, Y.~Z., \& {Wasserburg}, G.~J. 2007, \physrep, 442, 237,
  \dodoi{10.1016/j.physrep.2007.02.006}

\bibitem[{{Rauscher} {et~al.}(2002){Rauscher}, {Heger}, {Hoffman}, \&
  {Woosley}}]{rau02}
{Rauscher}, T., {Heger}, A., {Hoffman}, R.~D., \& {Woosley}, S.~E. 2002, \apj,
  576, 323, \dodoi{10.1086/341728}

\bibitem[{{Rickey} \& {Sheline}(1968)}]{rickey68}
{Rickey}, F.~A., \& {Sheline}, R.~K. 1968, Physical Review, 170, 1157,
  \dodoi{10.1103/PhysRev.170.1157}

\bibitem[{{Ritter} {et~al.}(2018{\natexlab{a}}){Ritter}, {C{\^o}t{\'e}},
  {Herwig}, {Navarro}, \& {Fryer}}]{rit18}
{Ritter}, C., {C{\^o}t{\'e}}, B., {Herwig}, F., {Navarro}, J.~F., \& {Fryer},
  C.~L. 2018{\natexlab{a}}, \apjs, 237, 42, \dodoi{10.3847/1538-4365/aad691}

\bibitem[{{Ritter} {et~al.}(2018{\natexlab{b}}){Ritter}, {Herwig}, {Jones},
  {Pignatari}, {Fryer}, \& {Hirschi}}]{nugrid2}
{Ritter}, C., {Herwig}, F., {Jones}, S., {et~al.} 2018{\natexlab{b}}, \mnras,
  480, 538, \dodoi{10.1093/mnras/sty1729}

\bibitem[{{Rizzuti} {et~al.}(2019){Rizzuti}, {Cescutti}, {Matteucci},
  {Chieffi}, {Hirschi}, \& {Limongi}}]{riz19}
{Rizzuti}, F., {Cescutti}, G., {Matteucci}, F., {et~al.} 2019, \mnras, 489,
  5244, \dodoi{10.1093/mnras/stz2505}

\bibitem[{{Sakuma} {et~al.}(2020){Sakuma}, {Hidaka}, \& {Yoneda}}]{sak20}
{Sakuma}, K., {Hidaka}, H., \& {Yoneda}, S. 2020, Geochemical Journal, 54, 393,
  \dodoi{10.2343/geochemj.2.0610}

\bibitem[{{Straniero} {et~al.}(2006){Straniero}, {Gallino}, \&
  {Cristallo}}]{str06}
{Straniero}, O., {Gallino}, R., \& {Cristallo}, S. 2006, \nphysa, 777, 311,
  \dodoi{10.1016/j.nuclphysa.2005.01.011}

\bibitem[{{Takahashi} \& {Yokoi}(1987)}]{tak87}
{Takahashi}, K., \& {Yokoi}, K. 1987, Atomic Data and Nuclear Data Tables, 36,
  375, \dodoi{10.1016/0092-640X(87)90010-6}

\bibitem[{{Thielemann} {et~al.}(2011){Thielemann}, {Arcones}, {K{\"a}ppeli},
  {Liebend{\"o}rfer}, {Rauscher}, {Winteler}, {Fr{\"o}hlich}, {Dillmann},
  {Fischer}, {Martinez-Pinedo}, {Langanke}, {Farouqi}, {Kratz}, {Panov}, \&
  {Korneev}}]{thi11}
{Thielemann}, F.~K., {Arcones}, A., {K{\"a}ppeli}, R., {et~al.} 2011, Progress
  in Particle and Nuclear Physics, 66, 346, \dodoi{10.1016/j.ppnp.2011.01.032}

\bibitem[{{Travaglio} {et~al.}(1999){Travaglio}, {Galli}, {Gallino}, {Busso},
  {Ferrini}, \& {Straniero}}]{trav99}
{Travaglio}, C., {Galli}, D., {Gallino}, R., {et~al.} 1999, \apj, 521, 691,
  \dodoi{10.1086/307571}

\bibitem[{{Travaglio} {et~al.}(2004){Travaglio}, {Gallino}, {Arnone}, {Cowan},
  {Jordan}, \& {Sneden}}]{trav04}
{Travaglio}, C., {Gallino}, R., {Arnone}, E., {et~al.} 2004, \apj, 601, 864,
  \dodoi{10.1086/380507}

\bibitem[{{Travaglio} {et~al.}(2001){Travaglio}, {Gallino}, {Busso}, \&
  {Gratton}}]{travaglio01}
{Travaglio}, C., {Gallino}, R., {Busso}, M., \& {Gratton}, R. 2001, \apj, 549,
  346, \dodoi{10.1086/319087}

\bibitem[{{Travaglio} {et~al.}(2015){Travaglio}, {Gallino}, {Rauscher},
  {R{\"o}pke}, \& {Hillebrandt}}]{travaglio:2015}
{Travaglio}, C., {Gallino}, R., {Rauscher}, T., {R{\"o}pke}, F.~K., \&
  {Hillebrandt}, W. 2015, \apj, 799, 54, \dodoi{10.1088/0004-637X/799/1/54}

\bibitem[{{Trippella} {et~al.}(2016){Trippella}, {Busso}, {Palmerini},
  {Maiorca}, \& {Nucci}}]{trip16}
{Trippella}, O., {Busso}, M., {Palmerini}, S., {Maiorca}, E., \& {Nucci}, M.~C.
  2016, \apj, 818, 125, \dodoi{10.3847/0004-637X/818/2/125}

\bibitem[{{Tsujimoto} {et~al.}(2017){Tsujimoto}, {Yokoyama}, \&
  {Bekki}}]{tsuj17}
{Tsujimoto}, T., {Yokoyama}, T., \& {Bekki}, K. 2017, \apjl, 835, L3,
  \dodoi{10.3847/2041-8213/835/1/L3}

\bibitem[{{Vescovi} {et~al.}(2018){Vescovi}, {Busso}, {Palmerini}, {Trippella},
  {Cristallo}, {Piersanti}, {Chieffi}, {Limongi}, {Hoppe}, \&
  {Kratz}}]{vesconi18}
{Vescovi}, D., {Busso}, M., {Palmerini}, S., {et~al.} 2018, \apj, 863, 115,
  \dodoi{10.3847/1538-4357/aad191}

\bibitem[{{Wagstaff} {et~al.}(2020){Wagstaff}, {Miller Bertolami}, \&
  {Weiss}}]{wag2020}
{Wagstaff}, G., {Miller Bertolami}, M.~M., \& {Weiss}, A. 2020, \mnras, 493,
  4748, \dodoi{10.1093/mnras/staa362}

\bibitem[{{Wasserburg} {et~al.}(2006){Wasserburg}, {Busso}, {Gallino}, \&
  {Nollett}}]{wasserburg06}
{Wasserburg}, G.~J., {Busso}, M., {Gallino}, R., \& {Nollett}, K.~M. 2006,
  \nphysa, 777, 5, \dodoi{10.1016/j.nuclphysa.2005.07.015}

\bibitem[{{Wasserburg} {et~al.}(1994){Wasserburg}, {Busso}, {Gallino}, \&
  {Raiteri}}]{wasserburg94}
{Wasserburg}, G.~J., {Busso}, M., {Gallino}, R., \& {Raiteri}, C.~M. 1994,
  \apj, 424, 412, \dodoi{10.1086/173899}

\bibitem[{{Wasserburg} {et~al.}(2017){Wasserburg}, {Karakas}, \&
  {Lugaro}}]{wasserburg17}
{Wasserburg}, G.~J., {Karakas}, A.~I., \& {Lugaro}, M. 2017, \apj, 836, 126,
  \dodoi{10.3847/1538-4357/836/1/126}

\bibitem[{{Wielen} {et~al.}(1996){Wielen}, {Fuchs}, \& {Dettbarn}}]{wielen1996}
{Wielen}, R., {Fuchs}, B., \& {Dettbarn}, C. 1996, \aap, 314, 438

\bibitem[{{Wu}(2005)}]{wu05}
{Wu}, S.~C. 2005, Nuclear Data Sheets, 106, 367,
  \dodoi{10.1016/j.nds.2005.11.001}

\bibitem[{{Yag{\"u}e L{\'o}pez} {et~al.}(2021){Yag{\"u}e L{\'o}pez},
  {C{\^o}t{\'e}}, \& {Lugaro}}]{yague21}
{Yag{\"u}e L{\'o}pez}, A., {C{\^o}t{\'e}}, B., \& {Lugaro}, M. 2021, \apj, 915,
  128, \dodoi{10.3847/1538-4357/ac02bf}

\bibitem[{{Yagüe López} {et~al.}(2021){Yagüe López}, {García-Hernández},
  {Ventura}, {Doherty}, {Hartogh}, {Jones}, \& {Lugaro}}]{snuppat}
{Yagüe López}, A., {García-Hernández}, A., {Ventura}, P., {et~al.} 2021,
  SNUPPAT, \url{https://github.com/AndresYague/Snuppat},  GitHub

\end{thebibliography}
\bibliographystyle{aasjournal}

\end{document}